\documentclass[%
aip,
 amsmath,amssymb,
preprint,%
linenumber
]{revtex4-2}

\usepackage{graphicx}
\usepackage{dcolumn}
\usepackage{bm}

\usepackage[utf8]{inputenc}
\usepackage[T1]{fontenc}
\usepackage{mathptmx}
\usepackage{etoolbox}
\usepackage{color}
\usepackage{dashrule}
\usepackage{caption}
\usepackage{subcaption}
\usepackage{textcmds}
\usepackage{placeins}
\usepackage{color}
\usepackage{xcolor}
\usepackage{hyperref}
\usepackage{epstopdf}

\newcommand\Rey{\mathit{Re}}
\newcommand\Pran{\mathit{Pr}}
\newcommand\Pen{\mathit{Pe}}
\makeatletter
\def\@email#1#2{%
 \endgroup
 \patchcmd{\titleblock@produce}
  {\frontmatter@RRAPformat}
  {\frontmatter@RRAPformat{\produce@RRAP{*#1\href{mailto:#2}{#2}}}\frontmatter@RRAPformat}
  {}{}
}%
\makeatother

\begin{document}

\title{Validation of symmetry-induced high moment velocity and temperature scaling laws in a turbulent channel flow}

\author{Francisco Alc\'antara-\'Avila}
\altaffiliation{Presently at: FLOW, Engineering Mechanics, KTH Royal Institute of Technology, 114 28 Stockholm, Sweden. Email: fraa@kth.se}
\author{Luis Miguel Garc\'ia-Raffi}
\author{Sergio Hoyas}
\affiliation{Instituto Universitario de Matem\'atica Pura y Aplicada, 
Universitat Polit\`ecnica de Val\`encia, Valencia 46022, Spain}

\author{Martin Oberlack}
\email{oberlack@fdy.tu-darmstadt.de}
\affiliation{Chair of Fluid Dynamics, TU Darmstadt, Otto-Bernd-Strasse 2, 64287 Darmstadt, Germany}
\altaffiliation{Also at: Centre for Computational Engineering, TU Darmstadt, Dolivostrasse 15, 64293 Darmstadt, Germany}

\begin{abstract}
The symmetry-based turbulence theory has been used to derive new scaling laws for the streamwise velocity and temperature moments of arbitrary order. For this, it has been applied to an incompressible turbulent channel flow driven by a pressure gradient with a passive scalar equation coupled in. To derive the scaling laws, symmetries of the classical Navier-Stokes and the thermal energy equations have been used together with statistical symmetries, i.e. the statistical scaling and translation symmetries of the multi-point moment equations. Specifically, the multi-point moments are built on the instantaneous velocity and temperature fields other than in the classical approach, where moments are based on the fluctuations of these fields. With this instantaneous approach, a linear system of multi-point correlation equations has been obtained, which greatly simplifies the symmetry analysis. The scaling laws have been derived in the limit of zero viscosity and heat conduction, i.e. $Re_\tau \rightarrow \infty$ and $Pr > 1$, and apply in the centre of the channel, i.e. they represent a generalization of the deficit law so herewith extending the work of Ref.~\onlinecite{obe22}. The scaling laws are all power laws, with the exponent of the high moments all depending exclusively on those of the first and second moments. To validate the new scaling laws, the data from a large number of DNS for different Reynolds and Prandtl numbers have been used. The results show a very high accuracy of the scaling laws to represent the DNS data. The statistical scaling symmetry of the multi-point moment equations, which characterizes intermittency, has been the key to the new results since it generates a constant in the exponent of the final scaling law. Most important, since this constant is independent of the order of the moments, it clearly indicates anomalous scaling.
\end{abstract}
\maketitle
\section{Introduction}\label{intro}

The open problem in physics with most applications in daily life is probably the behaviour of turbulent flows. Different strategies have been proposed when dealing with predicting turbulence in engineering. From the different approaches known in Computational Fluid Dynamics (CFD), Direct Numerical Simulations (DNS) has proven to be a powerful tool to generate highly reliable data bases for theoretical concepts on the nature of turbulence. In a DNS, no empirical modelling is needed to account for turbulent effects, and the approximations of the solutions of the Navier-Stokes equations are obtained through highly accurate numerical schemes. The main problematic issue of DNSs is their high computational cost since even the smallest scales of turbulence, the Kolmogorov scales, have to be simulated. Hence, this limits DNSs to very simple canonical geometries. However, DNSs have the same validity as experiments, and almost any imaginable quantity can be computed. 

It was not until the late 1980s that supercomputers could run the first DNS of turbulent flows. Specifically, in 1987, Kim et al. \cite{kim87} conducted the first DNS of a turbulent channel flow where a pressure gradient drove the flow at a low Reynolds number. The first DNS of a thermal turbulent channel flow was also performed in 1987 by Kim et al. \cite{kim87a}. The flow was also driven by a pressure gradient at a friction Reynolds number of $\Rey_\tau = 180$. Therein and presently, $\Rey_\tau$ is defined as $h u_\tau/\nu$, where $h$ is the semi-height of the channel, $u_\tau$ is the friction velocity, and $\nu$ is the kinematic viscosity of the fluid. The friction velocity is defined as $u_\tau = \sqrt{\tau_w/\rho}$, where $\tau_w$ is the averaged wall shear stress, and $\rho$ is the fluid density. Different Prandtl numbers were used, namely $\Pran = 0.1$, $0.71$ and $2$. Here the Prandtl number is defined as the ratio between the momentum diffusivity to the thermal diffusivity, $\Pran = \nu/\alpha$, where $\alpha$ is the thermal diffusivity of the fluid. One of the main results from the latter work was the validation of the DNS results comparing several first-order statistics with experimental data.

Since then, the aim of the DNS of thermal channel flows has been to increase the simulated Reynolds number, usually around $Pr = 0.71$, which is the Prandtl number of the air. However, reaching higher Reynolds numbers has a computational cost which scales as $L_x^2 L_z \Rey_\tau^4 \Pran^{3/2}$, according to Ref.~\onlinecite{yan99}. The largest DNS to date of a thermal channel flow\cite{pir16} used a friction Reynolds number of $4000$, and more recently, for isothermal turbulent channel flows, a friction Reynolds number of $Re_\tau = 10000$ was reached in Ref.~\onlinecite{hoy22}. However, these values are still far below the actual Reynolds numbers of most real-life problems. Therefore, Reynolds number extrapolations of the turbulent behaviour must be made, introducing inevitable errors and uncertainties. For example, the viscous diffusion and dissipation of the streamwise velocity fluctuations present a scaling failure near the wall \citep{hoy08}. An analogous scaling failure was obtained for the temperature variance at moderate Reynolds numbers. However, it was recently found \citep{alc21a} that for high Prandtl numbers and sufficiently high Reynolds numbers, the value of the viscous diffusion and dissipation of the temperature variance presented a much better scaling near the wall. Therefore, it is still an open question whether the streamwise velocity scaling failure will occur at much higher Reynolds numbers.

For all these reasons, turbulence is and will be, for many more years, an open problem without a complete analytical solution. Many researchers have proposed what are called turbulence scaling laws to describe the universal behaviour of turbulent statistics for special flows, though usually limited to the first and second moments. The most well-known scaling law is the universal law of the wall, which describes the profile of the mean streamwise velocity near the wall, consisting of a linear viscous sub-layer, where $\overline{U}^+ = y^+$, followed by the buffer layer and the logarithmic region further away from the wall. The overbar here denotes averaged in time, and the superscript $^+$ refers to dimensionless variables normalized in wall units with $u_\tau$, $\theta_\tau$ and $\nu$, where $\theta_\tau = q_w/(\rho c_p u_\tau)$ is the friction temperature, and $q_w$ and $c_p$ are the normal heat flux to the walls and the specific heat at constant pressure, respectively. Analogously to the law of the wall for the velocity, one can approximate the mean temperature near the wall in a similar form, where, for the first sub-layer, called conductive sub-layer, $\overline{\Theta}^+ = \Pran\hspace{0.1 cm}y^+$. In particular, the discovery of a logarithmic behaviour of the flow dates back to von K\'arm\'an in 1931. However, no connection between the analytical form of the scaling law and the Navier-Stokes equation was made. Despite the fact that a perfectly developed logarithmic region has not been observed in DNS due to the high Reynolds number needed to be simulated, a clear tendency suggests that it will appear for higher Reynolds numbers in different types of flows such as boundary layers, Couette flows, pipe flows, \cite{hoy06,jim08,avs14,pir14,lee15,loz14,kra18a,pir23}, etc. A first derivation of the logarithmic behaviour of the flow, based on first principles, was presented in Ref.~\onlinecite{obe01}. The use of Lie symmetries was the mathematical tool to achieve this.

Lie symmetries are a powerful mathematical theory to develop turbulent flow scaling laws. The origin of the Lie symmetries method dates back to the end of the 19th century when the mathematician Sophus Lie proposed it for obtaining solutions to differential equations and, most importantly, to systems of partial differential equations (PDE), such as the Navier-Stokes equations. The basis of the method consists of finding the symmetries of the system of PDE. Here, symmetry refers to a variable transformation that leads to an identical system of PDE, i.e. the transformed system of PDE has the same solution as the original one. With these symmetries, one can formulate a characteristic system (see \textsection\ref{sec3} for more details about the characteristic system), which in turn, leads to what is known as invariant solutions, which in turbulence are also known as scaling laws.

Lie symmetries possess several advantageous properties using ad hoc methods for a concrete application. First, symmetries can be obtained using computer algebra methods such as Maple. Second, symmetries give fundamental insight into the physics of the problem. And third, the scaling laws obtained are solutions to the moment equations and, hence, are based on first principles, not just pure curve fits. For these reasons, Lie symmetries are one of the most powerful tools for obtaining scaling laws of turbulent flows. Most important, it is also applicable to an infinite number of equations such as the moment hierarchy, and in this sense the ubiquitous closure problem of turbulence can be circumvented. This is also the approach that is presently applied.

The method has been widely studied by Oberlack and co-workers in several papers. Starting with Ref.~\onlinecite{obe01}, scaling laws for the three regions of wall-parallel shear flows (viscous sub-layer, logarithmic law, and deficit law in the centre of the channel) were obtained. Classical mechanical symmetries of the Navier-Stokes equations were used, but the key to the analysis was to employ the Multi-Point Correlation (MPC) equations. Two additional symmetries, not visible in the Navier-Stokes equations and called statistical symmetries, first discovered in Ref.~\onlinecite{obe10}, were used to derive the scaling laws that describe the flow statistics even for high moments. The next section will give more details about the MPC equations (\textsection\ref{sec2}). After this successful application of Lie symmetries to turbulent flows, several more works have been done with different geometries or boundary conditions \citep{avs14b,ros14,obe15,sad18,sad20a,sad20b}.

In this work, Lie symmetries theory will be used to derive new moment scaling laws of velocity and temperature, and mixed moments of arbitrary orders, extending the work in Ref.~\onlinecite{obe22} to include the temperature.  To achieve this, symmetries of the energy equation and the MPC equations of the energy and heat fluxes equations are obtained, from which the new scaling laws are formulated. These new scaling laws will be validated using the DNS data obtained by the authors in previous works (see \Citealp{alc18,alc21a,alc21b}). We will restrict ourselves to moments in the streamwise direction. However, this shows the ability of our method to obtain scaling laws following only strong mathematical arguments.  

In the next section, the governing equations of the problem are presented. In the third section, the Lie symmetries method is introduced, together with the application of the method to the governing equations of the problem. Then, in section four, the new scaling laws are developed and validated using DNS data. Finally, the fifth and last section contains conclusions.

\section{Governing equations}\label{sec2}

The equations that describe the behaviour of a turbulent flow, considering a Newtonian fluid with constant density and viscosity, are the well-known Navier-Stokes equations. For the sake of readability, the temporal and spatial dependencies will be omitted if uniqueness allows doing so. In the most general form, these equations can be written as
\begin{align}
\mathcal{C}(\boldsymbol{x}) &= \frac{\partial U_k}{\partial x_k} = 0, \label{continuity}\\
\mathcal{M}_i(\boldsymbol{x}) &= \frac{\partial U_i}{\partial t} + U_k\frac{\partial U_i}{\partial x_k} + \frac{\partial P}{\partial x_i} - \nu\frac{\partial^2 U_i}{\partial x_k \partial x_k} = 0, \label{momentum}
\end{align}
where $t \in \mathbb{R}^+$ is time; $x_i$ and $U_i$ are the space coordinate and velocity, $i=1,2,3$; and $P$ is the pressure divided by the density. The no-slip boundary condition is applied to both walls, periodic boundary conditions are used in the $x_1$ and $x_3$ directions and, to propel the flow, a constant pressure gradient is introduced in the $x_1$ direction so that the mass flux remains constant. The $x_2$ coordinate points in wall normal direction. Additionally, the thermal energy equation is simulated, which for a constant thermal conductivity coefficient, $\alpha$, reads
\begin{equation}
\mathcal{E}(\boldsymbol{x}) = \frac{\partial \Theta}{\partial t} + U_k\frac{\partial \Theta}{\partial x_k} - \alpha\frac{\partial^2 \Theta}{\partial x_k \partial x_k} = 0,
\label{energy}
\end{equation}
where $\Theta$ is the temperature. It should be noted that a constant heat flux at the wall was assumed in the simulation of the energy equation (\ref{energy}) since only this implies the temperature-scaling laws. This thermal boundary condition is known as the uniform heat flux (UHF) boundary condition. Curious reader is referred to Ref.~\onlinecite{kas92,alc18,alc21a,alc21b,llu21a} for more information about the UHF boundary condition. This is similar to the constant wall shear stress, which is central to the velocity-scaling laws. With this, since no heat sink was introduced, a constant temperature gradient in the $x_1$ direction is generated. This is removed from the flow by the following transformation to guarantee homogeneity in $x_1$-direction
\begin{equation}
\Theta = \langle\Theta_w\rangle_{x_3} - \Theta_{tr},
\label{equation_tranformed_temperature}
\end{equation}
where $\langle\Theta_w\rangle_{x_3}$ is the temperature at the wall averaged in time and in the $x_3$ direction, and $\Theta_{tr}$ is the transformed temperature. Therefore, $\langle\Theta_w\rangle_{x_3}$ carries the linear increment of the temperature and only depends on the $x_1$ direction. Then, $\Theta_{tr}$ is homogeneous in this streamwise direction. The somewhat unusual choice of the sign in (\ref{equation_tranformed_temperature}) is due to the fact that the temperature at the wall is maximum and therefore the transformed temperature $\Theta_{tr}$ remains positive throughout.

This allows the use of spectral discretization in the $x_1$ direction. Obviously, the scaling of $\Theta$ and $\Theta_{tr}$ in the $x_2$ direction will be the same since only a constant value will differ among them. Furthermore, since the scaling laws are presented as defect laws, for both $\Theta$ and $\Theta_{tr}$, these scaling laws must be the same. For the sake of generality, and without loss of veracity, we will refer to scaling laws of temperature, $\Theta$, instead of transformed temperature.

Using the Reynolds decomposition, one can separate the instantaneous variables (capital letter) in an average part (capital letter and over-bar) that does not depend on time, and a fluctuation part (lower case), e.g., $U_i(\boldsymbol{x},t) = \overline{U}_i(\boldsymbol{x}) + u_i(t,\boldsymbol{x})$ (note that temporal and spatial dependencies have been recovered only to show this example). Therefore, the following properties can be applied: the average in time of a mean quantity, $\overline{\Phi}$, will remain unchanged, i.e., $\overline{\overline{\Phi}} = \overline{\Phi}$; and the average in time of a fluctuation quantity is $0$, i.e., $\overline{\phi} = 0$. In addition, the following simplifications are valid for a developed turbulent channel flow driven by a pressure gradient,
\begin{align}
&\overline{U}_1 = \overline{U}_1(x_2),\quad \overline{P} = \overline{P}(x_1,x_2),\quad \overline{\Theta} = \overline{\Theta}(x_2), \nonumber\\
&\overline{U}_2 = \overline{U}_3 = 0,\quad \overline{u_i u_j} = \overline{u_i u_j}(x_2),\quad \overline{u_i \theta} = \overline{u_i \theta}(x_2).
\label{conditions}
\end{align}

Introducing the latter into equations (\ref{momentum}) and (\ref{energy}) and, in turn, taking the average, the governing equations reduce to
\begin{align}
\frac{d \overline{u_1 u_2}}{d x_2} + \frac{\partial \overline{P}}{\partial x_1} - \nu\frac{d^2 \overline{U}_1}{d x_2^2} &= 0, \label{mom_1_pro}\\
\frac{d \overline{u_2 u_2}}{d x_2} + \frac{\partial \overline{P}}{\partial x_2} &= 0, \label{mom_2_pro}\\
\frac{d \overline{u_3 u_2}}{d x_2} &= 0, \label{mom_3_pro}\\
\frac{d \overline{\theta u_2}}{d x_2}  - \alpha\frac{d^2 \overline{\Theta}}{d x_2^2} &= 0. \label{ener_pro}
\end{align}

Besides the latter one-point quantities, we may define the two-point correlation functions, or two-point moments, based on the fluctuating velocity,
\begin{equation}
R_{ij}(\boldsymbol{x},\boldsymbol{r}) = \overline{u_i(\boldsymbol{x}) u_j(\boldsymbol{x}+\boldsymbol{r})},\quad   R_{ij}^0(\boldsymbol{x}) = \lim_{\boldsymbol{r}\rightarrow 0} R_{ij}(\boldsymbol{x},\boldsymbol{r}) = \overline{u_i(\boldsymbol{x}) u_j(\boldsymbol{x})}.
\label{two_p_corr_ins}
\end{equation}

Employing an equivalent definition based on the instantaneous variables reads
\begin{equation}
H_{ij}(\boldsymbol{x},\boldsymbol{r}) = \overline{U_i(\boldsymbol{x}) U_j(\boldsymbol{x}+\boldsymbol{r})},\quad   H_{ij}^0(\boldsymbol{x}) = \lim_{\boldsymbol{r}\rightarrow 0} H_{ij}(\boldsymbol{x},\boldsymbol{r}) = \overline{U_i(\boldsymbol{x}) U_j(\boldsymbol{x})},
\label{def1}
\end{equation}
and a relation between the two correlation functions $R_{ij}$ and $H_{ij}$ reads as follows
\begin{equation}
R_{ij}(\boldsymbol{x},\boldsymbol{r}) = H_{ij}(\boldsymbol{x},\boldsymbol{r}) - \overline{U}_i(\boldsymbol{x})\overline{U}_j(\boldsymbol{x}+\boldsymbol{r}).
\end{equation}

This two-point concept can be extended for any number of points and ultimately forms the basis of the following analysis as well as the resulting scaling laws. Hence, we introduce the MPC equations (see e.g. Ref.~\onlinecite{obe15,sad18,sad20a,sad20b}). For high-order moments of velocity and temperature, they give additional information that is not provided in the one-point statistic equations, such as length scales. Also, when deriving a higher-order moment equation, only one unclosed function arises. As observed in equation (\ref{two_p_corr_ins}), from the two-point statistics one can obtain every one-point statistic. Finally, regarding Lie symmetries, two extra symmetries are obtained from the MPC equations, which are the key for determining the new scaling laws of the high-order moments, and will be pointed out in section \textsection\ref{sec4}.

Equations (\ref{two_p_corr_ins}) and (\ref{def1}) are the basis of the two different approaches that can be used to obtain the MPC equations: the fluctuating approach or the instantaneous approach. On one hand, the fluctuating approach has some advantages such as a straightforward relation to the Reynolds stress tensor or the turbulent heat fluxes. However, as noted in Ref.~\onlinecite{obe10}, a non-linear system of equations is obtained. Furthermore, all moment equations are coupled to the mean velocity or temperature, and equations of the third moment or higher, are coupled to the second moment. All this complicates the symmetry analysis that will be done below. On the other hand, the instantaneous approach results in a linear system of MPC equations with an equivalent but much simpler symmetry analysis. For this reason, the instantaneous approach is the one used in this work. It should be noted that the fluctuating approach and the instantaneous approach are bijective, i.e. mathematically physically absolutely equivalent.

Before presenting the MPC equations, some notations must be clarified. The correlation functions for the velocity are defined as
\begin{equation}
H_{i_{\{n\}}} = H_{i_{(1)}i_{(2)}\dots i_{(n)}} = \overline{U_{i_{(1)}}(\boldsymbol{x}_{(1)}) U_{i_{(2)}}(\boldsymbol{x}_{(2)})\dots U_{i_{(n)}}(\boldsymbol{x}_{(n)})},
\label{def_U}
\end{equation}
which for $n = 2$, $\boldsymbol{x}_{(1)} = \boldsymbol{x}$ and $\boldsymbol{x}_{(2)} = \boldsymbol{x} + \boldsymbol{r}$ yields to (\ref{def1}). Note the in $U_{i_{(n)}}(\boldsymbol{x}_{(n)})$, the subscript $i_{(n)}$ refers to the velocity direction of the $n$-th term which is measured at the coordinate $\boldsymbol{x}_{(n)}$. The definition of the temperature correlation is
\begin{equation}
H_{\Theta_{\{m\}}} = H_{\Theta_{(1)}\Theta_{(2)}\dots \Theta_{(m)}} = \overline{\Theta(\boldsymbol{x}_{(1)}) \Theta(\boldsymbol{x}_{(2)})\dots \Theta(\boldsymbol{x}_{(m)})}.
\label{def_T}
\end{equation}

Mixed moments of velocity and temperature, which in the limit of only one temperature and one velocity reduces to the turbulent heat flux, reads
\begin{align}
& H_{i_{\{n\}}\Theta_{\{m\}}} = H_{i_{(1)}i_{(2)}\dots i_{(n)}\Theta_{(n+1)}\Theta_{(n+2)}\dots \Theta_{(n+m)}} = \nonumber\\
& \overline{U_{i_{(1)}}(\boldsymbol{x}_{(1)}) U_{i_{(2)}}(\boldsymbol{x}_{(2)})\dots U_{i_{(n)}}(\boldsymbol{x}_{(n)})\Theta(\boldsymbol{x}_{(n+1)}) \Theta(\boldsymbol{x}_{(n+2)})\dots \Theta(\boldsymbol{x}_{(n+m)})}.
\label{def_HF}
\end{align}

Note that (\ref{def_U}) and (\ref{def_T}) are just particular cases of (\ref{def_HF}) for $m$ and $n$ equal to $0$, respectively, but for the sake of readability, they are presented separately. When pressure is involved in the correlation, the notation, in the general form, is
\begin{align}
&I_{i_{\{n-1\}}\Theta_{\{m\}}[l]_{P}} = H_{i_{(1)}\dots i_{(l-1)}Pi_{(l+1)}\dots i_{(n)}\Theta_{(n+1)}\Theta_{(n+2)}\dots \Theta_{(n+m)}} = \nonumber\\
&\overline{U_{i_{(1)}}(\boldsymbol{x}_{(1)})\dots  P(\boldsymbol{x}_{(l)}) \dots  U_{i_{(n)}}(\boldsymbol{x}_{(n)})\Theta(\boldsymbol{x}_{(n+1)})\dots \Theta(\boldsymbol{x}_{(n+m)})},
\label{def_PP}
\end{align}
for $1\le l\le n$. Finally, the following notation
\begin{align}
&H_{i_{\{n\}}\Theta_{\{m\}}[i_{(l)}\rightarrow k]}(\boldsymbol{x}_{(l)}\rightarrow\boldsymbol{x}_{(p)}) = \nonumber\\
&\overline{U_{i_{(1)}}(\boldsymbol{x}_{(1)})\dots  U_{i_{(l-1)}}(\boldsymbol{x}_{(l-1)})U_k(\boldsymbol{x}_{(p)})U_{i_{(l+1)}}(\boldsymbol{x}_{(l+1)}) \dots  U_{i_{(n)}}(\boldsymbol{x}_{(n)})\Theta(\boldsymbol{x}_{(n+1)})\dots \Theta(\boldsymbol{x}_{(n+m)})}
\label{def_change}
\end{align}
is used to indicate a change in the correlation function of velocity direction, $i_{(l)}$, to $k$ and/or the coordinate where the variable is applied, $\boldsymbol{x}_{(l)}$, to $\boldsymbol{x}_{(p)}$. With these definitions, the MPC equations of the heat flux moments of order $n+m$ reads (see Appendix \ref{appA} for detailed step-by-step derivation of the MPC equations)
\begin{align}
& \frac{\partial H_{i_{\{n\}}\Theta_{\{m\}}}}{\partial t} \nonumber\\
& + \sum_{a=1}^n\left(\frac{\partial H_{i_{\{n+1\}}\Theta_{\{m\}}[i_{(n+m+1)}\rightarrow k]}(\boldsymbol{x}_{(n+m+1)}\rightarrow\boldsymbol{x}_{(a)})}{\partial x_{k_{(a)}}} + \frac{\partial I_{i_{\{n-1\}}\Theta_{\{m\}}[a]_{P}}}{\partial x_{i_{(a)}}} - \nu\frac{\partial^2 H_{i_{\{n\}}\Theta_{\{m\}}}}{\partial x_{k_{(a)}}\partial x_{k_{(a)}}}\right) \nonumber\\
& + \sum_{b=n+1}^{n+m}\left(\frac{\partial H_{i_{\{n+1\}}\Theta_{\{m\}}[i_{(n+m+1)}\rightarrow k]}(\boldsymbol{x}_{(n+m+1)}\rightarrow\boldsymbol{x}_{(b)})}{\partial x_{k_{(b)}}} - \alpha\frac{\partial^2 H_{i_{\{n\}}\Theta_{\{m\}}}}{\partial x_{k_{(b)}}\partial x_{k_{(b)}}}\right) = 0.
\label{MPC}
\end{align}

As was mentioned before in (\ref{def_HF}), the MPC equations of the velocity and temperature moments are specific cases of (\ref{MPC}), which can be obtained by setting, respectively, $m$ and $n$ equals to $0$.

Additionally, the continuity equations read
\begin{align}
&\frac{\partial H_{i_{\{n\}}\Theta_{\{m\}}[i_{(l)}\rightarrow k]}}{\partial x_{k_{(l)}}} = 0 \quad \mbox{for}\quad l=1,2,\dots ,n, \label{MPC_C1}\\
&\frac{\partial I_{i_{\{n-1\}}\Theta_{\{m\}}[a]_P[i_{(l)}\rightarrow k]}}{\partial x_{k_{(l)}}} = 0 \quad \mbox{for}\quad a,l=1,2,\dots ,n, \quad a\neq l,\quad \mbox{and}\quad n \ge 2.
\label{MPC_C2}
\end{align}

Note that pure temperature correlations and heat fluxes correlations with $l > n$ do not admit continuity equations since they would have originated from $\partial \overline{\Theta}(\boldsymbol{x})/\partial x_k$, which is not a continuity equation.

As was previously noted, the system of the MPC equations (\ref{MPC}), (\ref{MPC_C1}), and (\ref{MPC_C2}) is linear for any turbulent flow. Moreover, the dependent variables $H$ and $I$ appear inside spatial or temporal derivatives. As seen in sections \textsection\ref{sec3_3} and \textsection\ref{sec4}, this is the key to obtaining two important statistical Lie symmetries necessary to derive the scaling laws.

To make the notation easier to understand, the Two-Point Correlation (TPC) equations for the velocity, heat fluxes, and temperature are presented in Appendix \ref{appB}.

\section{Lie symmetries of the MPC equations}\label{sec3}

In this section, the Lie symmetries method will be presented briefly. After that, the symmetries of the governing equations introduced in the previous section will be given.

\subsection{Symmetry transformations}

Given a system of partial differential equations (PDE)
\begin{equation}
F(\boldsymbol{x},\boldsymbol{y},\boldsymbol{y}^{(1)},\boldsymbol{y}^{(2)},\dots ) = 0,
\label{PDE}
\end{equation}
where $\boldsymbol{x}$ are the independent variables, $\boldsymbol{y}$ are the dependent variables, and $\boldsymbol{y}^{(n)}$ are the $n$-th derivative of the dependent variables with respect to all coordinate combinations of $\boldsymbol{x}$. Based on this, a transformation of (\ref{PDE}), with the form
\begin{equation}
\boldsymbol{x}^* = \boldsymbol{\phi}(\boldsymbol{x},\boldsymbol{y}),\quad \boldsymbol{y}^* = \boldsymbol{\psi}(\boldsymbol{x},\boldsymbol{y}),
\label{symmetry}
\end{equation}
is called a symmetry transformation, or just symmetry, if the following holds:
\begin{equation}
\boldsymbol{F}(\boldsymbol{x},\boldsymbol{y},\boldsymbol{y}^{(1)},\boldsymbol{y}^{(2)},\dots ) = 0 \hspace{0.3cm}\Leftrightarrow\hspace{0.3cm} \boldsymbol{F}(\boldsymbol{x}^*,\boldsymbol{y}^*,\boldsymbol{y}^{*(1)},\boldsymbol{y}^{*(2)},\dots ) = 0.
\end{equation}

In other words, a symmetry transformation (\ref{symmetry}) leaves the PDEs (\ref{PDE}) invariant and, in addition, maps any solution of (\ref{PDE}) into a new solution.

In Lie group analysis, it is the aim to find all possible symmetry transformations of the PDE (\ref{PDE}). The notation group refers to the fact that symmetry transformations usually admit group properties. As we are presently dealing with Lie symmetry groups, the group parameter $\varepsilon \in \mathbb{R}$ has to be introduced to obtain the so-called one-parameter Lie symmetry group of transformations, with the form
\begin{equation}
\boldsymbol{x}^* = \boldsymbol{\phi}(\boldsymbol{x},\boldsymbol{y};\varepsilon),\quad \boldsymbol{y}^* = \boldsymbol{\psi}(\boldsymbol{x},\boldsymbol{y};\varepsilon).
\label{Lie_symmetry}
\end{equation}

Equation (\ref{Lie_symmetry}) provides a continuous group of transformations that allows analytical solutions for the underlying equations. As for the group properties of (\ref{Lie_symmetry}), we may, without loss of generality, assign $\varepsilon = 0$ to the identity element, i.e.
\begin{equation}
\boldsymbol{x}^* = \boldsymbol{\phi}(\boldsymbol{x},\boldsymbol{y};\varepsilon=0) = \boldsymbol{x},\quad \boldsymbol{y}^* = \boldsymbol{\psi}(\boldsymbol{x},\boldsymbol{y};\varepsilon=0) = \boldsymbol{y}.
\label{Lie_symmetry_ident}
\end{equation}

Therefore, if a Taylor series at $\varepsilon = 0$ is applied to the Lie group of transformation  (\ref{Lie_symmetry}), we obtain
\begin{align}
\boldsymbol{x}^* = \boldsymbol{x} + \left.\frac{\partial \boldsymbol{\phi}(\boldsymbol{x},\boldsymbol{y};\varepsilon)}{\partial \varepsilon}\right|_{\varepsilon = 0}\varepsilon + O(\varepsilon^2) = \boldsymbol{x} + \boldsymbol{\xi}(\boldsymbol{x},\boldsymbol{y})\varepsilon + O(\varepsilon^2), \label{symmetrie_infinites_x}\\
\boldsymbol{y}^* = \boldsymbol{y} + \left.\frac{\partial \boldsymbol{\psi}(\boldsymbol{x},\boldsymbol{y};\varepsilon)}{\partial \varepsilon}\right|_{\varepsilon = 0}\varepsilon + O(\varepsilon^2) = \boldsymbol{y} + \boldsymbol{\eta}(\boldsymbol{x},\boldsymbol{y})\varepsilon + O(\varepsilon^2). \label{symmetrie_infinites_y}
\end{align}

Equations (\ref{symmetrie_infinites_x}) and (\ref{symmetrie_infinites_y}) are the infinitesimal form of the Lie group of transformation (\ref{Lie_symmetry}), where $\boldsymbol{\xi}$ and $\boldsymbol{\eta}$ are the so-called infinitesimals. The general form of the transformation (\ref{Lie_symmetry}) and the infinitesimal transformations (\ref{symmetrie_infinites_x}) and (\ref{symmetrie_infinites_y}) are related by Lie's first theorem (see Ref.~\onlinecite{blu02}), i.e. if the infinitesimals of the transformation, $\boldsymbol{\xi}$ and $\boldsymbol{\eta}$ are known, one can uniquely recover the general form of the symmetry group of transformations (\ref{Lie_symmetry}). In order to obtain all symmetries of a given system of PDEs, one has to invoke the infinitesimal form of the transformations (\ref{symmetrie_infinites_x}) and (\ref{symmetrie_infinites_y}). Based on this, an algebraic algorithm, the Lie algorithm, evolves, which has been implemented into various computer algebra systems. Details on the algorithm may be taken from different textbooks like Ref.~\onlinecite{blu02} or in works such as Ref.~\onlinecite{ros14}.

\subsection{Group invariant solutions}

In addition to the fact that symmetries characterize fundamental physical properties of a system, it is the ability to construct solutions that are central to their application, and which will also be applied here. For this purpose, we define the so-called group-invariant solutions, i.e. once the Lie symmetries of the system of PDE are obtained, the next step is to generate invariant solutions, which in turbulence are referred to as scaling laws. We call $\boldsymbol{y} = \boldsymbol{\Psi}(\boldsymbol{x})$ an invariant solution of a PDE system if and only if

\begin{enumerate}
\item $\boldsymbol{y} - \boldsymbol{\Psi}(\boldsymbol{x})$ is invariant under $X$, where $X$ is the so-called infinitesimal generator, defined as
\begin{equation}
X = \xi_i(\boldsymbol{x},\boldsymbol{y})\frac{\partial}{\partial x_i} + \eta_j(\boldsymbol{x},\boldsymbol{y})\frac{\partial}{\partial y_j}.
\label{inf_gen}
\end{equation}

Hence, we have
\begin{equation}
X(\boldsymbol{y} - \boldsymbol{\Psi}(\boldsymbol{x})) = 0,
\end{equation}
on $\boldsymbol{y} = \boldsymbol{\Psi}(\boldsymbol{x})$. Using the operator (\ref{inf_gen}) and differentiating out, we obtain the following hyperbolic system
\begin{equation}
\xi_i(\boldsymbol{x},\boldsymbol{\Psi})\frac{\partial \Psi_j}{\partial \xi_i} = \eta_j(\boldsymbol{x},\boldsymbol{\Psi}),\quad i = 1,\dots ,k;\quad j = 1,\dots ,l,
\label{inv_sur_con}
\end{equation}
which generates the invariant solution. The solution of the hyperbolic system (\ref{inv_sur_con}) can now be determined by the method of characteristics and we obtain the so-called invariant surface condition

\begin{equation}
\frac{dx_1}{\xi_1} = \frac{dx_2}{\xi_2} = \dots  = \frac{dx_k}{\xi_k} = \frac{dy_1}{\eta_1} = \frac{dy_2}{\eta_2} = \dots  = \frac{dy_l}{\eta_l},
\label{inv_sur_con2}
\end{equation}
where $k$ and $l$ are the respective numbers of the independent and dependent variables.

The integrals of the system (\ref{inv_sur_con2}) are the characteristics of the hyperbolic system (\ref{inv_sur_con}) but, at the same time, the invariants of the original system of equations. These now form the basis of the invariant solutions and, thus, the new independent and dependent variables - the similarity variables.

\item Finally, the invariant solution, $\boldsymbol{y} = \boldsymbol{\Psi}(\boldsymbol{x})$ has to solve the PDE system, which is to be verified by insertion into the original PDE.
\end{enumerate}

\subsection{Symmetries of the governing equations}\label{sec3_3}

In this section, we present the Lie symmetries of the governing equations (\ref{continuity})-(\ref{energy}). For high Reynolds numbers flows, i.e., for $\Rey_\tau \rightarrow \infty$, and sufficiently away from the wall, the viscous effects are limited to length-scales of the order of the Kolmogorov length scale. In Oberlack\cite{obe00b}, this fact forms the basis for a singular asymptotic expansion similar to boundary layer theory. As a result, two sets of moment equations arise, where the equations for the "outer" solutions is frictionless and acts on length scales larger than the Kolmogorov scale, while an "inner" equation contains friction terms and operates on the Kolmogorov length. As a result, the frictionless "outer" equation and the corresponding solutions have the symmetries of the Euler equation. The above development again illustrates the fact that although the limit $\nu\to0^+$ can be considered, this is not identical to $\nu=0$. For the following analyses, this means a focus on the large scales and thus that $\nu=\alpha=0$ may be set in equation (\ref{MPC}), assuming $\nu\sim\alpha$, i.e. the diffusion terms are of the same order of magnitude. Incidentally, the dissipation therefore results from the "inner" equation, which is not considered presently.

For the case of the Euler equations, a 10-parameter symmetry group of transformation is obtained, where we here only present the scaling groups needed further below,
\begin{align}
T_{Sx}&:\quad t^* = t,\quad \boldsymbol{x}^* = \mathrm{e}^{a_{Sx}}\boldsymbol{x},\quad \boldsymbol{U}^* = \mathrm{e}^{a_{Sx}}\boldsymbol{U},\quad P^* = \mathrm{e}^{2a_{Sx}}P, \label{sym_sca_s}\\
T_{St}&:\quad t^* = \mathrm{e}^{a_{St}}t,\quad \boldsymbol{x}^* = \boldsymbol{x},\quad \boldsymbol{U}^* = \mathrm{e}^{-a_{St}}\boldsymbol{U},\quad P^* = \mathrm{e}^{-2a_{St}}P, \label{sym_sca_t}.
\end{align}

The coefficients $a_{Sx}$ and $a_{St}$ are the group parameters of scaling of space and time, respectively.

If the Navier-Stokes equations are considered, i.e., the viscous term is not neglected, the two scaling symmetries, $T_{Sx}$ and $T_{St}$, linearly combine into a simple scaling symmetry. This phenomenon, in which a multi-parameter symmetry group of transformations is reduced after a specific condition is applied, is known as symmetry breaking.

An analogous simplification, as the transition from the Navier-Stokes to the Euler equation, applied to the energy equation (\ref{energy}), can be done by neglecting the diffusive term, i.e., $\Pen_\tau \rightarrow \infty$, which holds in the center of the channel. Considering this, the energy equation admits the following infinite-dimensional symmetry,
\begin{equation}
T_{\Theta}:\quad t^* = t,\quad \boldsymbol{x}^* = \boldsymbol{x},\quad \boldsymbol{U}^* = \boldsymbol{U},\quad P^* = P,\quad \Theta^* = f\left(\Theta\right).
\label{sym_sca_theta}
\end{equation}

For scaling purposes, and in analogy with the scaling symmetries of the Euler equations, we consider the simplification $f(\Theta) = \mathrm{e}^{a_\Theta}\Theta$, so that $T_{\Theta}$ represents a scaling of temperature. It should be noted that the energy equation (\ref{energy}), just like the Navier-Stokes equations (\ref{continuity}) and (\ref{momentum}), admits the Galilean group.

As noted in Ref.~\onlinecite{obe10}, the symmetries obtained for the Navier-Stokes and energy equations transfer to the MPC equations (\ref{MPC}). So, in the limit of zero viscosity and diffusion, i.e. $Re_\tau \rightarrow \infty$ and $Pr > 1$, the MPC equations (\ref{MPC}) admit the following scaling symmetries:
\begin{align}
\overline{T}_{Sx}&:\quad t^* = t,\quad \boldsymbol{x}^* = \mathrm{e}^{a_{Sx}}\boldsymbol{x},\quad H_{i_{\{n\}}\Theta_{\{m\}}}^* = \mathrm{e}^{na_{Sx}}H_{i_{\{n\}}\Theta_{\{m\}}}, \nonumber\\
&\quad\enskip I_{i_{\{n-1\}}\Theta_{\{m\}}[a]_{P}}^* = \mathrm{e}^{(n+1)a_{Sx}}I_{i_{\{n-1\}}\Theta_{\{m\}}[a]_{P}}, \label{sym_MPC_sca_sp}\\
\overline{T}_{St}&:\quad t^* = \mathrm{e}^{a_{St}}t,\quad \boldsymbol{x}^* = \boldsymbol{x},\quad H_{i_{\{n\}}\Theta_{\{m\}}}^* = \mathrm{e}^{-na_{St}}H_{i_{\{n\}}\Theta_{\{m\}}}, \nonumber\\
&\quad\enskip I_{i_{\{n-1\}}\Theta_{\{m\}}[a]_{P}}^* = \mathrm{e}^{-(n+1)a_{St}}I_{i_{\{n-1\}}\Theta_{\{m\}}[a]_{P}}, \label{sym_MPC_sca_t}\\
\overline{T}_{S\Theta}&:\quad t^* = t,\quad \boldsymbol{x}^* = \boldsymbol{x},\quad H_{i_{\{n\}}\Theta_{\{m\}}}^* = \mathrm{e}^{ma_\Theta}H_{i_{\{n\}}\Theta_{\{m\}}}, \nonumber\\
&\quad\enskip I_{i_{\{n-1\}}\Theta_{\{m\}}[a]_{P}}^* = \mathrm{e}^{ma_\Theta}I_{i_{\{n-1\}}\Theta_{\{m\}}[a]_{P}}, \label{sym_MPC_sca_theta}
\end{align}
which are immediate consequences of (\ref{sym_sca_s}), (\ref{sym_sca_t}) and the scaling version of (\ref{sym_sca_theta}).

In addition to the symmetries induced from the Navier-Stokes/Euler and energy equations, the MPC equations (\ref{MPC}) admit an extended set of symmetry transformations. These symmetries are called statistical symmetries and they are the key in the process of deriving scaling laws \citep{obe15} for high-order moments of the velocity and temperature. These symmetries were discovered in Ref.~\onlinecite{obe10} and detailed information on the physical meaning of the statistical symmetries can be found in Ref.~\onlinecite{wac14}. First, because of the linearity of the MPC equations (\ref{MPC}), a scaling symmetry of the dependent variables is admitted
\begin{align}
\overline{T}_{Ss}&:\quad t^* = t,\quad \boldsymbol{x}^* = \boldsymbol{x},\quad H_{i_{\{n\}}\Theta_{\{m\}}}^* = \mathrm{e}^{a_{Ss}}H_{i_{\{n\}}\Theta_{\{m\}}}, \nonumber\\
&\quad\enskip I_{i_{\{n-1\}}\Theta_{\{m\}}[a]_{P}}^* = \mathrm{e}^{a_{Ss}}I_{i_{\{n-1\}}\Theta_{\{m\}}[a]_{P}}. \label{sym_MPC_sca}
\end{align}

This symmetry, as proven in Ref.~\onlinecite{wac14}, represents a measure of intermittency. For intermittency, we understand a flow with subsequently changing turbulent and non-turbulent regimes. Moreover, as all dependent variables in (\ref{MPC}) appear inside derivatives, a translation symmetry of all moments is also admitted
\begin{align}
\overline{T}_{tra,H}&:\quad t^* = t,\quad \boldsymbol{x}^* = \boldsymbol{x},\quad H_{i_{\{n\}}\Theta_{\{m\}}}^* = H_{i_{\{n\}}\Theta_{\{m\}}} + \boldsymbol{a}_{i_{\{n\}}\Theta_{\{m\}}}^H, \nonumber\\
&\quad\enskip I_{i_{\{n-1\}}\Theta_{\{m\}}[a]_{P}}^* = I_{i_{\{n-1\}}\Theta_{\{m\}}[a]_{P}} + \boldsymbol{a}_{i_{\{n-1\}}\Theta_{\{m\}}}^I. \label{sym_MPC_tra_H}
\end{align}

Apart from the symmetries presented, we will also include the classical translation in space symmetry, i.e.
\begin{align}
\overline{T}_{tra,x}&:\quad t^* = t,\quad \boldsymbol{x}^* = \boldsymbol{x} + \boldsymbol{a}_x,\quad H_{i_{\{n\}}\Theta_{\{m\}}}^* = H_{i_{\{n\}}\Theta_{\{m\}}}, \nonumber\\
&\quad\enskip I_{i_{\{n-1\}}\Theta_{\{m\}}[a]_{P}}^* = I_{i_{\{n-1\}}\Theta_{\{m\}}[a]_{P}}. \label{sym_MPC_tra_x}
\end{align}

Note that, in contrast to (\ref{sym_MPC_sca}), where $a_{Ss}$ is a single group parameter, symmetries (\ref{sym_MPC_tra_H}) and (\ref{sym_MPC_tra_x}) are a condensed way of showing several symmetries. Each component of the vector and tensors $\boldsymbol{a}_{\{n\}\{m\}}^H$, $\boldsymbol{a}_{\{n-1\}\{m\}}^I$ and $\boldsymbol{a}_x$ represent the group parameter of different and independent symmetries. Therefore, infinite symmetries are contained in (\ref{sym_MPC_tra_H}), while (\ref{sym_MPC_tra_x}) contains three symmetries, one for each spatial direction.

In summary, six symmetries have been identified, that will be used to derive the scaling laws of high-order moments of velocity and temperature. One property of the Lie symmetries is that one can combine different one-parameter Lie symmetries into a multi-parameter Lie symmetry. Following this, one can obtain the following multi-parameter Lie symmetry group from the symmetries (\ref{sym_MPC_sca_sp})-(\ref{sym_MPC_tra_x})
\begin{align}
T&:\quad t^* = \mathrm{e}^{a_{St}}t,\quad \boldsymbol{x}^* = \mathrm{e}^{a_{Sx}}\boldsymbol{x} + \boldsymbol{a}_x, \nonumber\\
&\quad\enskip H_{i_{\{n\}}\Theta_{\{m\}}}^* = \mathrm{e}^{n(a_{Sx}-a_{St})+ma_\Theta + a_{Ss}}H_{i_{\{n\}}\Theta_{\{m\}}} + \boldsymbol{a}_{i_{\{n\}}\Theta_{\{m\}}}^H, \label{sym_MPC}\\
&\quad\enskip I_{i_{\{n-1\}}\Theta_{\{m\}}[a]_{P}}^* = \mathrm{e}^{(n+1)(a_{Sx}-a_{St})+ma_\Theta + a_{Ss}}I_{i_{\{n-1\}}\Theta_{\{m\}}[a]_{P}} + \boldsymbol{a}_{i_{\{n-1\}}\Theta_{\{m\}}}^I. \nonumber
\end{align}

A different way of writing the symmetry group (\ref{sym_MPC}) is using the infinitesimal notation (\ref{symmetrie_infinites_x}) and (\ref{symmetrie_infinites_y}), from which one obtains
\begin{align}
&\xi_t = a_{St} t,\quad \boldsymbol{\xi}_x = a_{Sx} \boldsymbol{x} + \boldsymbol{a}_x, \nonumber\\
&\eta_{H_{i_{\{n\}}\Theta_{\{m\}}}} = \left[n(a_{Sx}-a_{St})+ma_\Theta + a_{Ss}\right]H_{i_{\{n\}}\Theta_{\{m\}}} + \boldsymbol{a}_{i_{\{n\}}\Theta_{\{m\}}}^H, \label{sym_MPC_inf}\\
&\eta_{I_{i_{\{n-1\}}\Theta_{\{m\}}[a]_{P}}} = \left[(n+1)(a_{Sx}-a_{St})+ma_\Theta + a_{Ss}\right]I_{i_{\{n-1\}}\Theta_{\{m\}}[a]_{P}} + \boldsymbol{a}_{i_{\{n-1\}}\Theta_{\{m\}}}^I, \nonumber
\end{align}
and which will be used below in the next chapter.

\section{High order moment scaling laws and its validation}\label{sec4}

\subsection{New velocity, temperature and mixed moment scaling laws}

Using the symmetries (\ref{sym_MPC}), or rather its infinitesimal form (\ref{sym_MPC_inf}), we are now able to compute the invariant solutions of the equations (\ref{MPC}), which in turbulence are called the turbulent scaling laws. It should be noted here that in the following we only consider the moments of the instantaneous variables, i.e. the $H$ approach. Of course, a conversion for each moment into that of the fluctuations is straighforward, but in Oberlack et al.\cite{obe23} we were able to show that an error accumulation occurs in the calculation from DNS data, which inevitably results from the finite number of available flow fields. This error increases considerably as the order of the moments increases. We have therefore deliberately refrained from displaying the moments from the fluctuations.

Further, we note that the study focuses now on the streamwise velocity and temperature. For this purpose, we only have to insert the infinitesimals (\ref{sym_MPC_inf}) into the invariant surface condition (\ref{inv_sur_con2}) and we get
\begin{align}
\frac{d x_2}{a_{Sx} x_2 + a_{x_2}} &= \frac{d H_{1_{\{1\}}}}{\left[a_{Sx}-a_{St} + a_{Ss}\right]H_{1_{\{1\}}} + \boldsymbol{a}_{1_{\{1\}}}^H} \nonumber\\
&= \frac{d H_{\Theta_{\{1\}}}}{\left[a_\Theta + a_{Ss}\right]H_{\Theta_{\{1\}}} + \boldsymbol{a}_{\Theta_{\{1\}}}^H} \nonumber\\
&= \frac{d H_{1_{\{1\}}\Theta_{\{1\}}}}{\left[a_{Sx}-a_{St} + a_\Theta + a_{Ss}\right]H_{1_{\{1\}}\Theta_{\{1\}}} + \boldsymbol{a}_{1_{\{1\}}\Theta_{\{1\}}}^H} \nonumber\\
&= \dots  \nonumber\\
&= \frac{d H_{1_{\{n\}}\Theta_{\{m\}}}}{\left[n(a_{Sx}-a_{St})+ma_\Theta + a_{Ss}\right]H_{1_{\{n\}}\Theta_{\{m\}}} + \boldsymbol{a}_{1_{\{n\}}\Theta_{\{m\}}}^H} \label{char_sys_pro}
\end{align}

Since we consider a shear flow that is fully developed in $x_1$ and $x_3$, all moments in the one-point limit depend only on $x_2$. Furthermore, the dependencies of the other points for the higher-order tensors would have to be formally considered as well, because this would result in further similarity variables. However, from now on, we will focus on one-point statistics, so that every point of application of the variables in (\ref{MPC}) will be $\boldsymbol{x}_{(1)}=\boldsymbol{x}_{(2)}=\dots =\boldsymbol{x}_{(n+m)}$. Integrating (\ref{char_sys_pro}), notice that we are using the first and the last term because the other terms are just specific moments, we obtain the following invariant solutions for any arbitrary moment
\begin{align}
H_{1_{\{n\}}\Theta_{\{m\}}} &= C'_{1_{\{n\}}\Theta_{\{m\}}}\left(x_2 + \frac{a_{x_2}}{a_{Sx}}\right)^{n(\sigma_2 - \sigma_1) + m\sigma_\Theta + 2\sigma_1 - \sigma_2} - \frac{\boldsymbol{a}_{1_{\{n\}}\Theta_{\{m\}}}^H}{n(a_{Sx}-a_{St})+m a_\Theta + a_{Ss}}, \nonumber\\
&\label{inv_sol}\\
&\rm{with}\quad C'_{1_{\{n\}}\Theta_{\{m\}}} = e^{c'_{nm}\left[n(a_{Sx} - a_{St}) + ma_\Theta + a_{Ss}\right]},\label{origin_prefactor}
\end{align}
where $c'_{nm}$ denote the constants of integration, $\sigma_1 = 1 - a_{St}/a_{Sx} + a_{Ss}/a_{Sx}$, $\sigma_2 = 2(1-a_{St}/a_{Sx}) + a_{Ss}/a_{Sx}$ and $\sigma_\Theta = a_\Theta/a_{Sx}$. Similar to Ref.~\onlinecite{obe22}, the choice of parameters in the exponent of (\ref{inv_sol}) has been designed so that the high-order moments depend on those of the first and second order. Focusing on the velocity only, as in Ref.~\onlinecite{obe22}, i.e. $m = 0$, the exponent for $n=1$ is $\sigma_1$, and for $n=2$, it is $\sigma_2$, while for pure temperature moments, i.e. $n = 0$, we have for $m=1$ the exponent $\sigma_\Theta$. Therefore, $\sigma_1$ and $\sigma_2$, are determined from the first two velocity moments, while $\sigma_\Theta$ is determined from the first temperature moment.

At this point, it is important to recall that the invariant solution (\ref{inv_sol}) has been derived in the limit of vanishing viscosity and heat conduction. Therefore, this solution will be only valid in the region where these conditions apply, i.e. the centre of the channel. The invariant solution (\ref{inv_sol}) shows that the moments of velocity, temperature, and higher-order moments scale as power-laws, whose exponents are determined by the parameters $\sigma_1$, $\sigma_2$ and $\sigma_\Theta$. Note that, in the exponent of the power law, four initial parameters appear ($a_{Sx}$, $a_{St}$, $a_\Theta$, and $a_{Ss}$). However, these parameters appear as ratios, leading to only three free parameters remaining ($\sigma_1$, $\sigma_2$, and $\sigma_\Theta$).

Analogously to Ref.~\onlinecite{obe01}, where the scaling law of the mean velocity of a turbulent shear flow was presented as a deficit law, and significantly extended in Ref.~\onlinecite{obe22} to arbitrary velocity moments, equation (\ref{inv_sol}) can be rewritten to form the final deficit scaling law of velocity, temperature and arbitrarily mixed moments of both as
\begin{align}
\frac{H_{{1_{\{n\}}\Theta_{\{m\}}}_{cl}} - H_{1_{\{n\}}\Theta_{\{m\}}}}{u_\tau^n \theta_\tau^m} &= C'_{nm} \left(\frac{x_2}{h}\right)^{n(\sigma_2 - \sigma_1) + m\sigma_\Theta + 2\sigma_1 - \sigma_2}, \label{scaling_law}\\
&\rm{with}\quad C'_{nm} = \alpha'e^{n\beta' + m\beta'_\Theta}, \label{prefactor}
\end{align}
where the subscript $cl$ refers to the value of the moment on the centre line, which comes from the last term on the right-hand side of the equation (\ref{inv_sol}), and $C'_{nm}$ are the new exponential scaling factors. Similar to Ref.~\onlinecite{obe22}, it has been assumed that $c'_{nm}$ is independent of $n$ and $m$ to derive $C'_{nm}$ in (\ref{scaling_law}), and this will indeed be validated below in section \textsection\ref{sec4_2}. Also note that, in equation (\ref{scaling_law}), the shift in $x_2$ has been set to $0$, as the coordinate is anchored to the centre line.

\subsection{Validation of the scaling law (\ref{scaling_law}) with DNS data}\label{sec4_2}

The new scaling law (\ref{scaling_law}) will be validated by using DNS data of turbulent channel flows driven by a pressure gradient at friction Reynolds numbers of $\Rey_\tau = 500$, $1000$, $2000$ and $5000$, and heated by a constant heat flux from both walls with a wide range of values of Prandtl numbers: $\Pran = 0.007$, $0.01$, $0.02$, $0.05$, $0.1$, $0.3$, $0.5$, $0.71$, $1$, $2$, $4$, $7$ and $10$. Specifically, the combinations of $\Rey_\tau$ and $\Pran$ used are presented in table \ref{sims}.

\begin{table}
\centering
\begin{tabular}{ccccccccccccccc}
$\Rey_\tau\backslash\Pran$ & \textbf{$0.007$} & \textbf{$0.01$} & \textbf{$0.02$} & \textbf{$0.05$} & \textbf{$0.1$} & \textbf{$0.3$} & \textbf{$0.5$} & \textbf{$0.71$} & \textbf{$1$} & \textbf{$2$} & \textbf{$4$} & \textbf{$7$} & \textbf{$10$} & Colour\\ \hline
\textbf{$500$}                     & X                & X               & X               & X               & X              & X              & X              & X               & X            & X            & X            & X            & X & \color{red}\hdashrule[0.55ex]{0.8cm}{2pt}{}            \\
\textbf{$1000$}                    & X                & X               & X               & X               & X              & X              & X              & X               & X            & X            & X            & X            &     & \color{green}\hdashrule[0.55ex]{0.8cm}{2pt}{}          \\
\textbf{$2000$}                    & X                & X               & X               & X               & X              & X              & X              & X               & X            & X            & X            & X            &     & \color{blue}\hdashrule[0.55ex]{0.8cm}{2pt}{}          \\
\textbf{$5000$}                    &                  &                 &                 &                 &                &                &                & X               &              &              &              &              &    & \color{black}\hdashrule[0.55ex]{0.8cm}{2pt}{}          
\end{tabular}
\caption{$\Rey_\tau$ and $\Pran$ numbers used for the validation of the scaling law (\ref{scaling_law}). The last column shows the colours used in the figures to refer to each Reynolds number.}
\label{sims}
\end{table}

The code used to run the simulations is the Liso code, already validated and employed in many other simulations of turbulent channel flows \citep{hoy06,hoy08,avs14,avs14b,gan18,llu18}. Detailed information about the code itself and the parameters of the simulation (mesh size, wash-outs run, computational box size,\dots ) can be found in Ref.~\onlinecite{alc18,alc21a,alc21b,llu21a}. As mentioned, the UHF is used as the thermal boundary condition in the DNS cited. This implies that temperature increases linearly in the streamwise direction. To make the temperature field homogeneous in the $x_1$ direction, the value of the temperature at the wall is removed, obtaining a transformed temperature. Because we want to give a general scaling law for temperature moments, and because the same symmetry, $T_{\Theta}$ (\ref{sym_sca_theta}), is also obtained for the energy equation of the transformed temperature, the general form of the temperature energy equation is used in this work.

The moments calculated in the simulations are limited to order seven for pure moments of velocity and temperature and six for mixed moments of velocity and temperature. The procedure to fit the scaling law (\ref{scaling_law}) to the DNS data has been done by minimizing the infinite norm of the relative error between the fit and the value of the DNS data, i.e.
\begin{equation}
\rm{error} = \min\left(\left|\left|\frac{data(x_2)-fit(x_2)}{data(x_2)}\right|\right|_\infty \right).
\label{fit}
\end{equation}

The infinite norm is used in this formula since the data takes values across several orders of magnitude. After this fitting is applied to the first and second moments of velocity and the first moment of temperature, $\sigma_1$, $\sigma_2$ and $\sigma_\Theta$ from the scaling law (\ref{scaling_law}) are determined and, thus, the exponent for any high order moment is known and only the constants of integration, $C'_{nm}$, must be calculated.

\begin{figure}
\centering
\begin{subfigure}[b] {0.49\textwidth}
\includegraphics[width=0.99\textwidth]{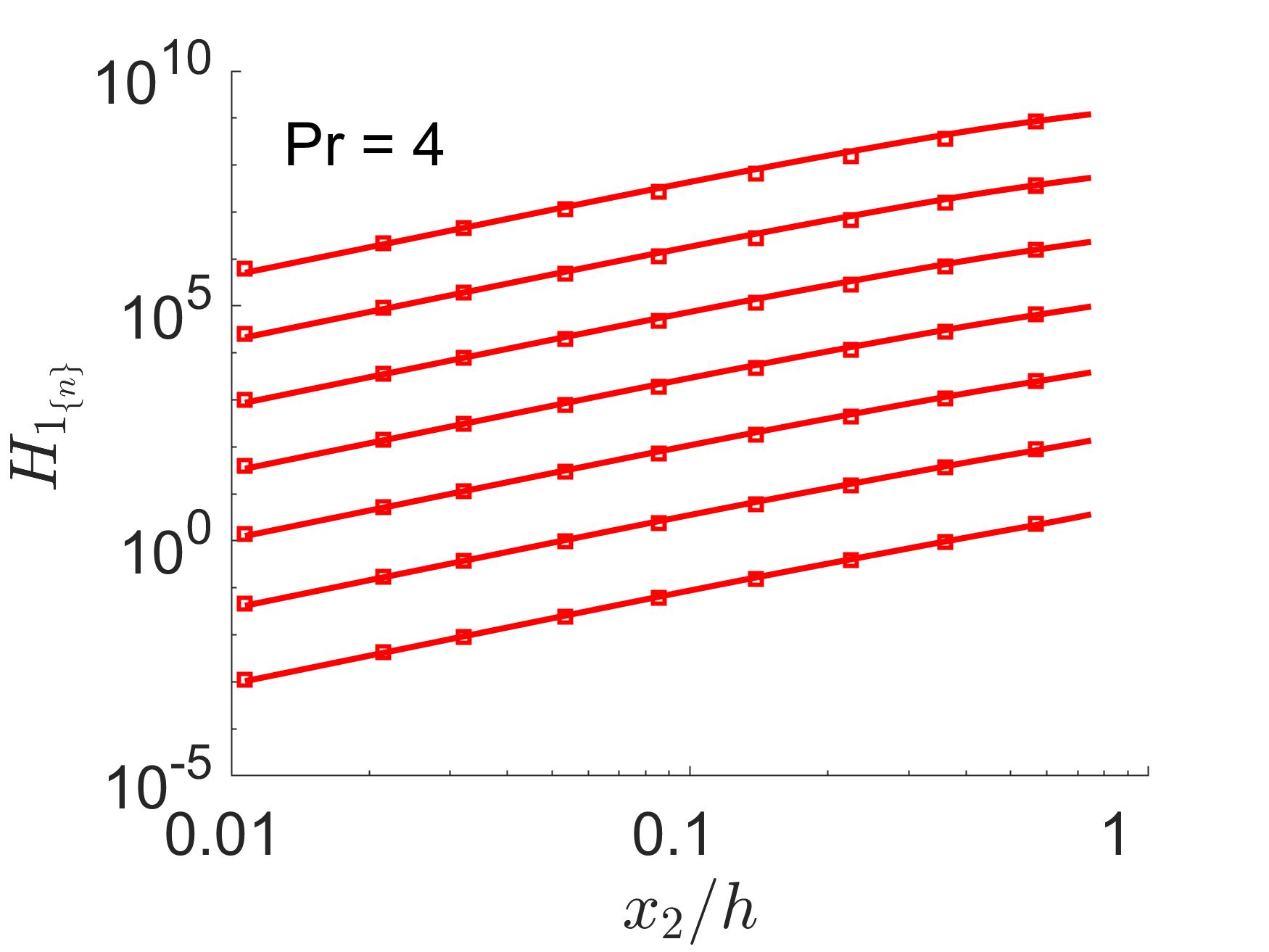}
	\caption{}
	\label{u_mom_fit_500_4}
\end{subfigure}
\centering
\begin{subfigure}[b] {0.49\textwidth}
\includegraphics[width=0.99\textwidth]{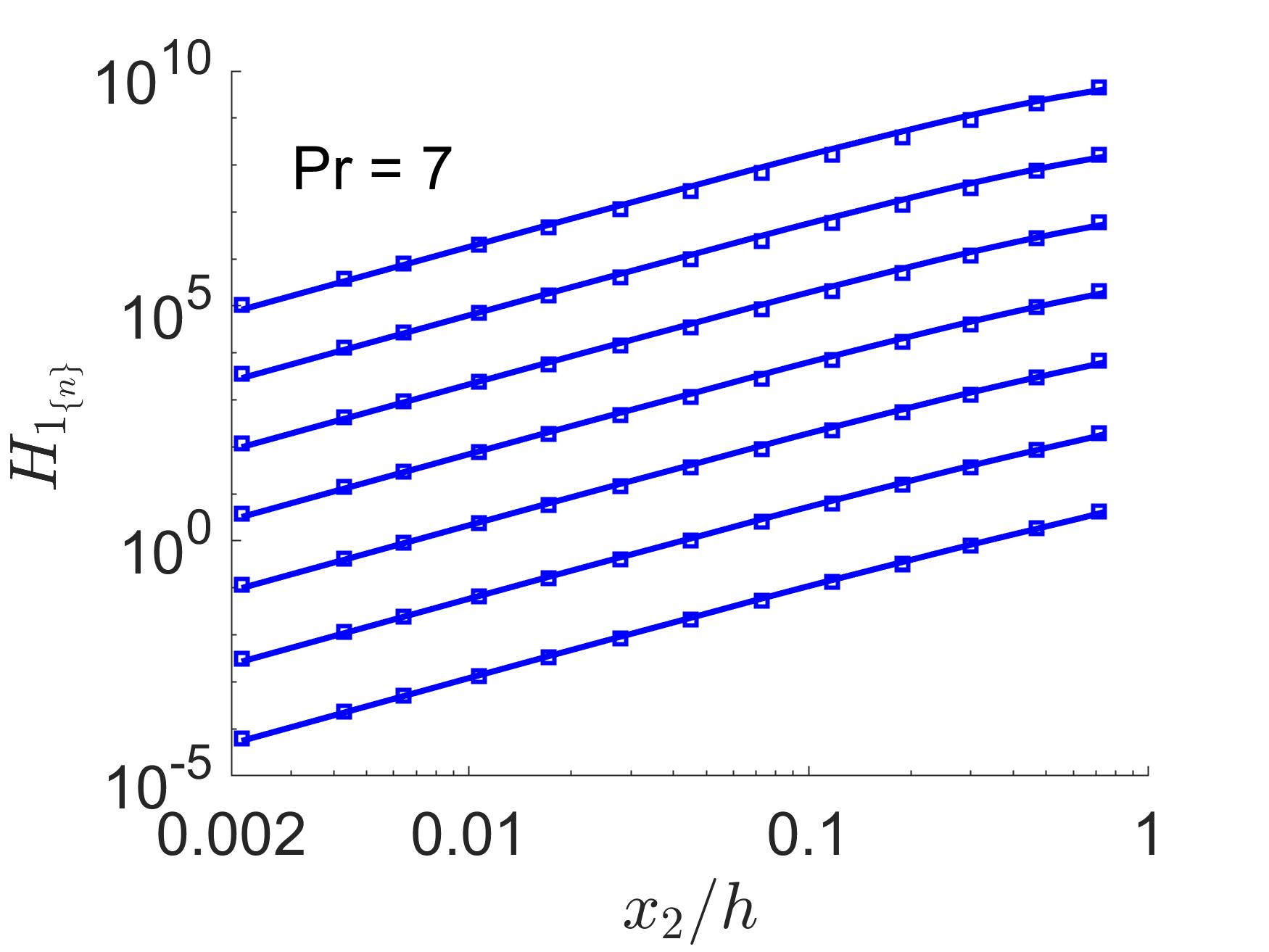}
	\caption{}
	\label{u_mom_fit_2000_7}
\end{subfigure}
\centering
\begin{subfigure}[b] {0.49\textwidth}
\includegraphics[width=0.99\textwidth]{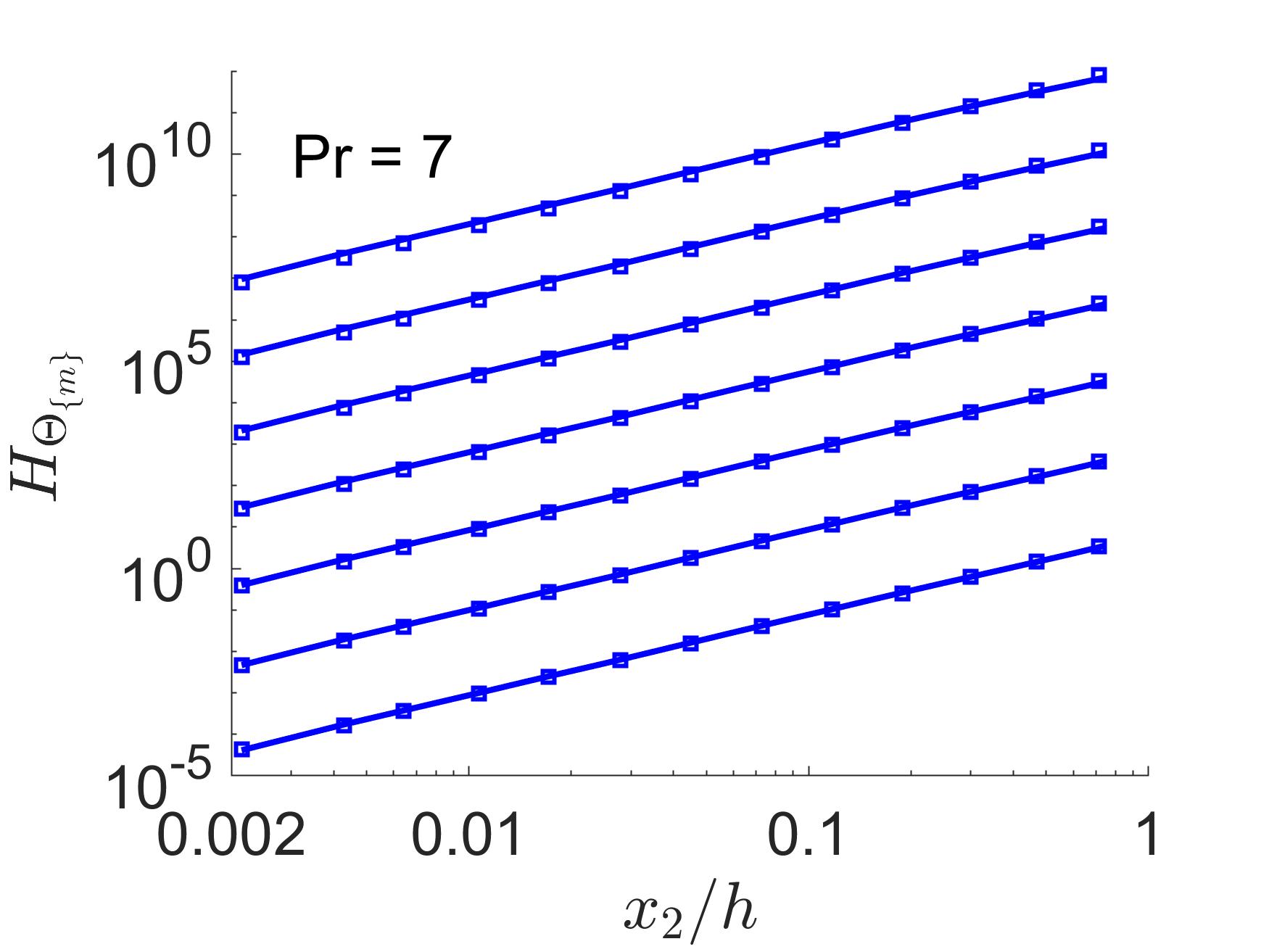}
	\caption{}
	\label{t_mom_fit_2000_7}
\end{subfigure}
\centering
\begin{subfigure}[b] {0.49\textwidth}
\includegraphics[width=0.99\textwidth]{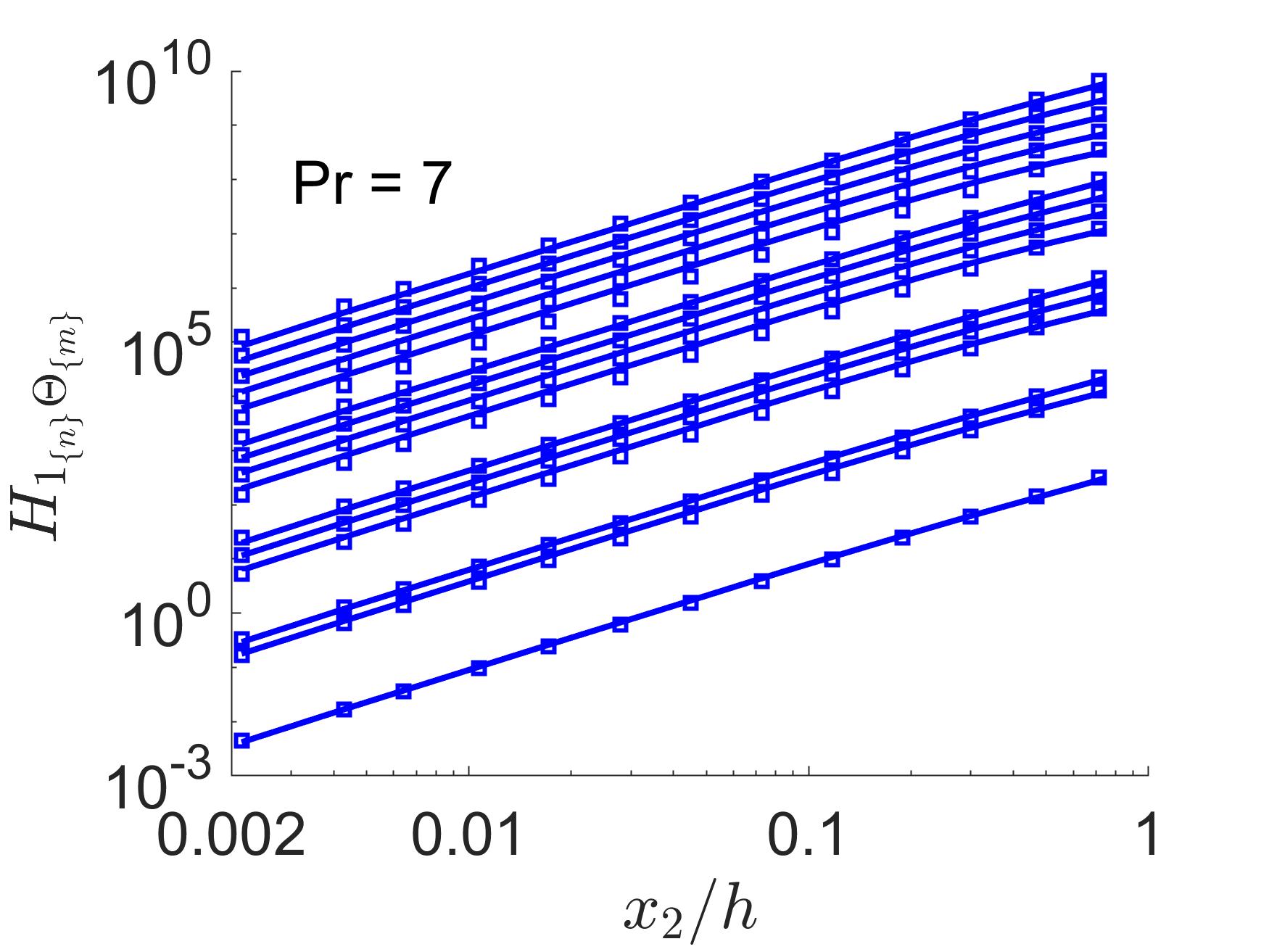}
	\caption{}
	\label{ut_mom_fit_2000_7}
\end{subfigure}
\caption{Moments of velocity, $H_{1_{\{n\}}}$, for (a) $\Rey_\tau = 500$ and $Pr = 4$ and (b) $\Rey_\tau = 2000$ and $Pr = 7$. Moments of (c) temperature, $H_{\Theta_{\{m\}}}$, and (d) heat fluxes, $H_{1_{\{n\}}\Theta_{\{m\}}}$, for $\Rey_\tau = 2000$ and $Pr = 7$. In (a), (b), and (c), velocity and temperature moments are obtained for $n$ and $m = 1$, $2$,\dots , $7$, appearing in that order from bottom to top of the plot. For (d), heat fluxes moments are shown for $n+m = 2$, $3$,\dots , $6$, appearing in that order from bottom to top of the plot. For heat fluxes moments of the same order, the lower lines are for $m = 0$, while the upper lines are for $n = 0$. Solid lines are the values from the DNS, while squares represent the values from the scaling law. The wall and centre of the channel are swapped, so the centre line is at $x_2/h = 0$, while the wall is at $x_2/h = 1$. Colours as in table \ref{sims}.}
\label{mom_fit_high}
\end{figure}

The results of the fits of the velocity moments for $\Rey_\tau = 500$ and $\Pran = 4$ are depicted in figure \ref{u_mom_fit_500_4}, together with the fits of the velocity moments for $\Rey_\tau = 2000$ and $\Pran = 7$, in figure \ref{u_mom_fit_2000_7}. Additionally, for $\Rey_\tau = 2000$ and $\Pran = 7$, the fits of the temperature moments and mixed moments are presented in figures \ref{t_mom_fit_2000_7} and \ref{ut_mom_fit_2000_7}, respectively. Solid lines represent the values from the DNS, while squares are the values obtained from the scaling law (\ref{scaling_law}). Recall that: in all figures below, the wall and centre of the channel are swapped, so the centre line is at $x_2/h = 0$, while the wall is at $x_2/h = 1$. Also, it is important to mention that the range of the centre of the channel where the scaling law has been applied is up to $x_2/h = 0.75$. The most important result of this work is the high accuracy of the scaling law (\ref{scaling_law}) to fit the data of the DNS for all moments, with the highest relative errors of only $0.2\%$ for the higher order moments, calculated with equation (\ref{fit}). Even for the lowest Reynolds numbers of value $500$, the accuracy of the fit is almost as good as for $\Rey_\tau = 2000$, as can be seen in figures \ref{u_mom_fit_500_4} and \ref{u_mom_fit_2000_7}. In the same way, the scaling law is validated with the same accuracy for the temperature and mixed moments as shown in figures \ref{t_mom_fit_2000_7} and \ref{ut_mom_fit_2000_7}, respectively.

\begin{figure}
\centering
\begin{subfigure}[b] {0.49\textwidth}
\includegraphics[width=0.99\textwidth]{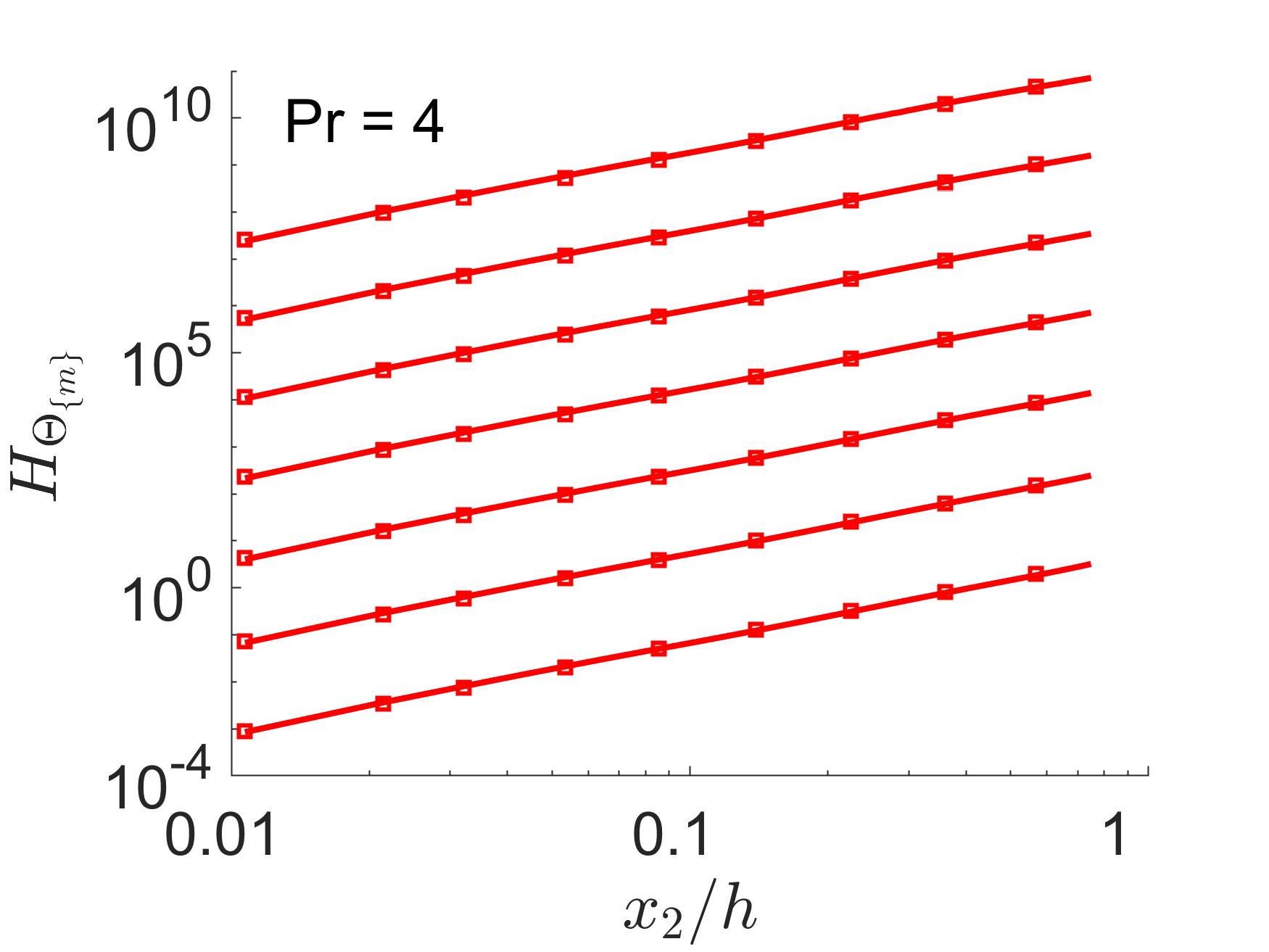}
	\caption{}
	\label{t_mom_fit_500_4}
\end{subfigure}
\centering
\begin{subfigure}[b] {0.49\textwidth}
\includegraphics[width=0.99\textwidth]{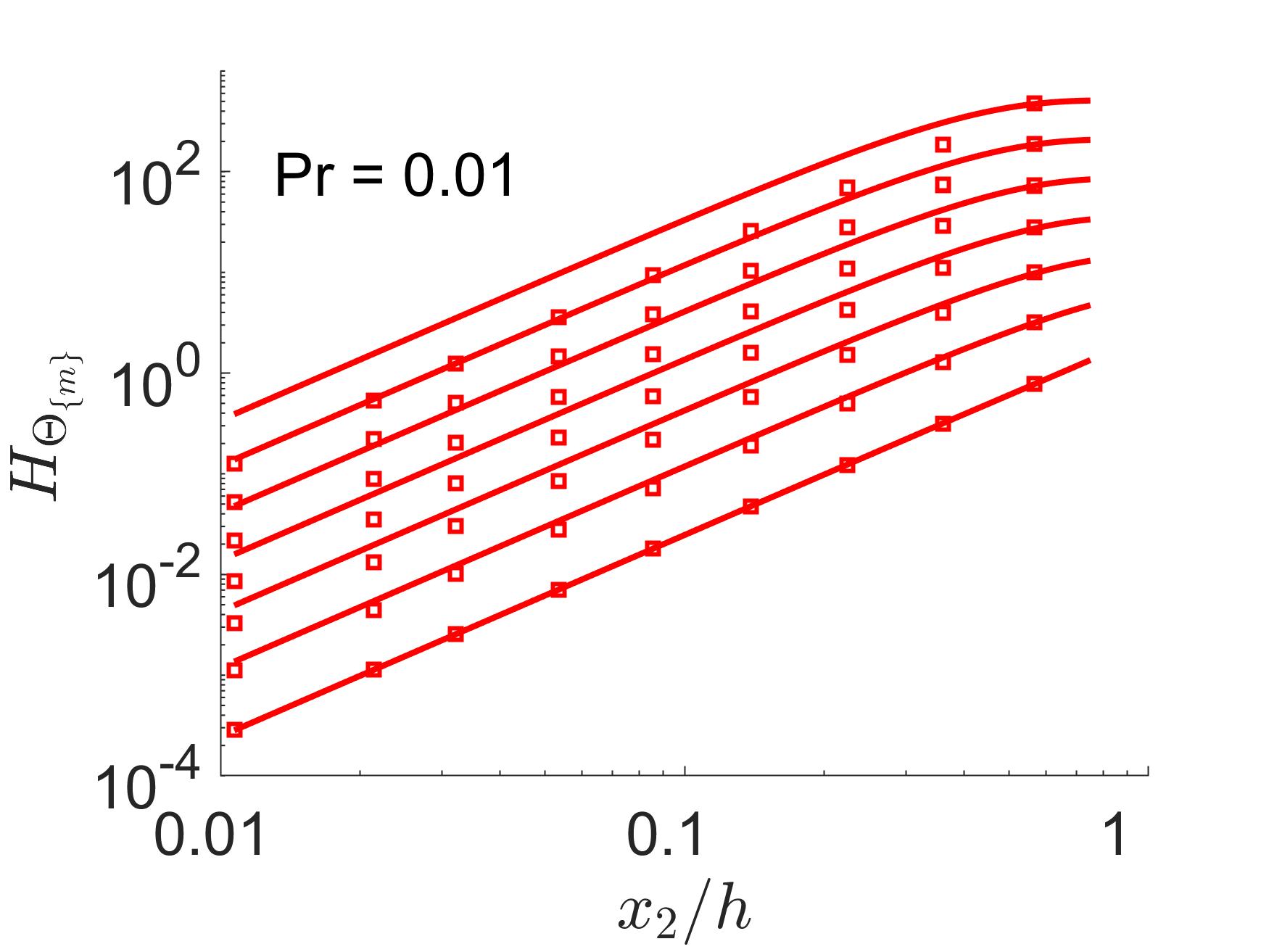}
	\caption{}
	\label{t_mom_fit_500_001}
\end{subfigure}
\centering
\begin{subfigure}[b] {0.49\textwidth}
\includegraphics[width=0.99\textwidth]{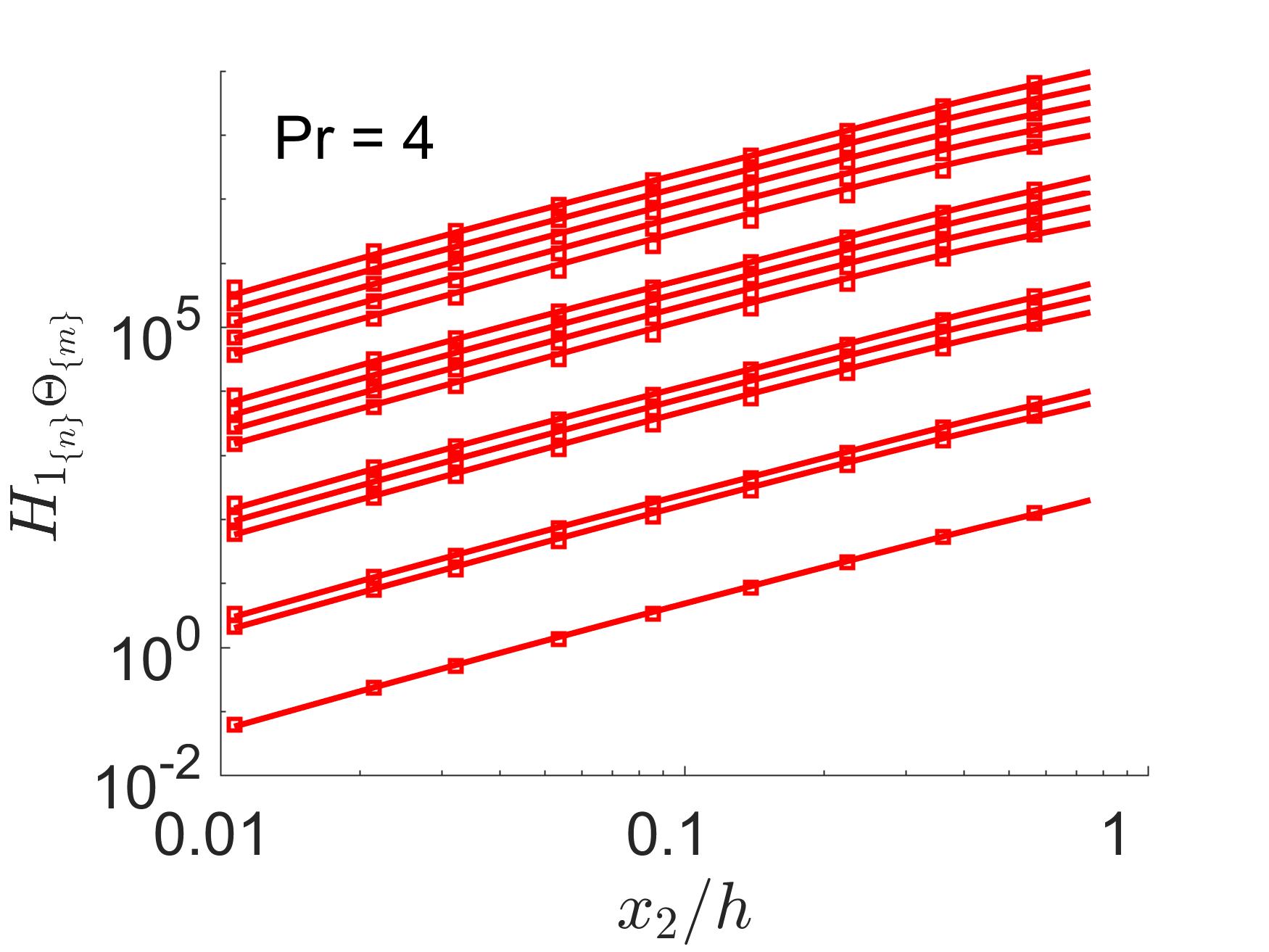}
	\caption{}
	\label{ut_mom_fit_500_4}
\end{subfigure}
\centering
\begin{subfigure}[b] {0.49\textwidth}
\includegraphics[width=0.99\textwidth]{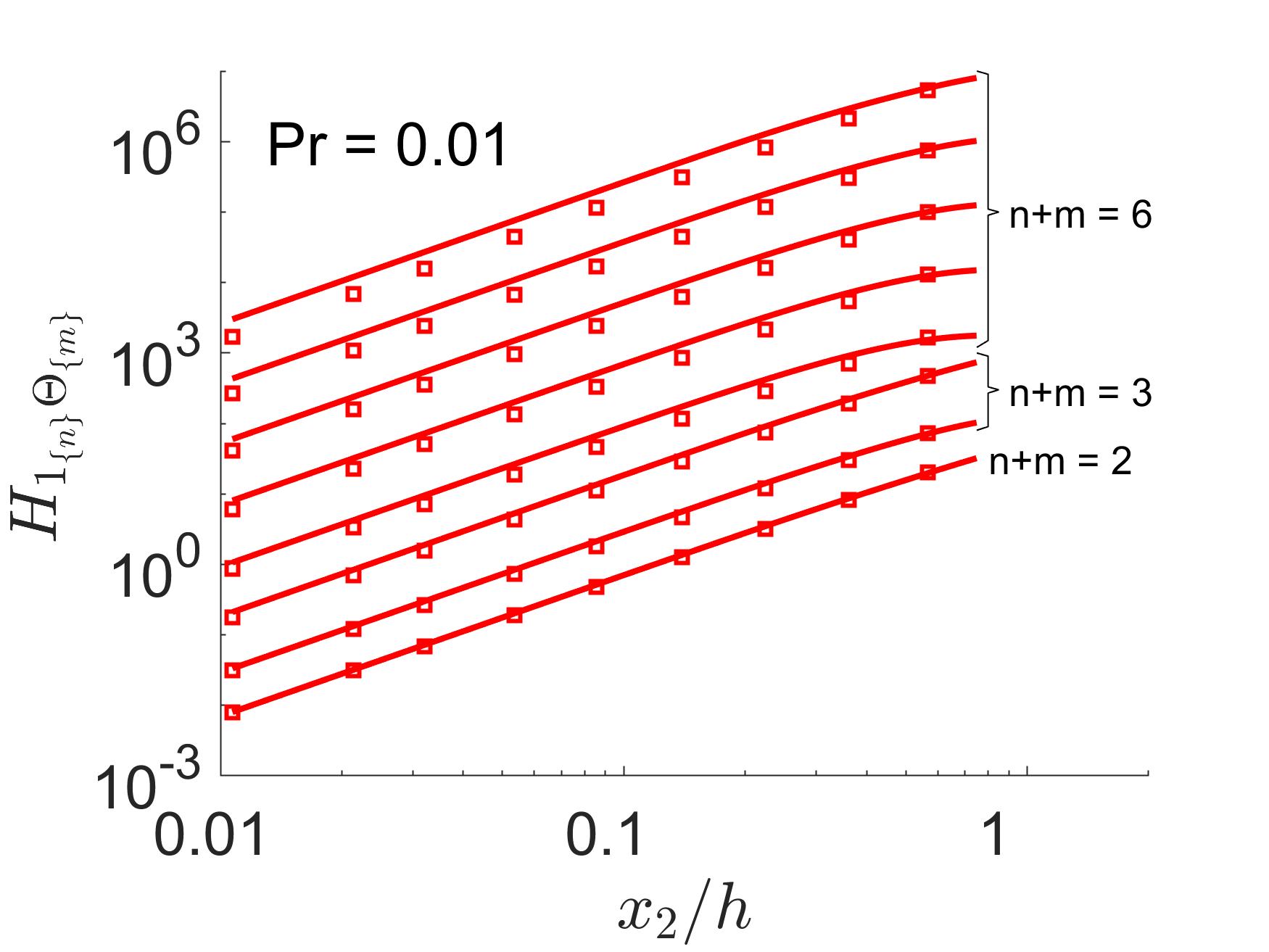}
	\caption{}
	\label{ut_mom_fit_500_001}
\end{subfigure}
\caption{Moments of temperature, $H_{\Theta_{\{m\}}}$, for (a) $\Rey_\tau = 500$ and $Pr = 4$ and (b) $\Rey_\tau = 500$ and $Pr = 0.01$. Moments of heat fluxes, $H_{1_{\{n\}}\Theta_{\{m\}}}$, for (c) $\Rey_\tau = 500$ and $Pr = 4$ and (d) $\Rey_\tau = 500$ and $Pr = 0.01$. In (a) and (b), temperature moments are obtained for $m = 1$, $2$,\dots , $7$, appearing in that order from the bottom to the top of the plot. For (c), heat fluxes moments are shown for $n+m = 2$, $3$,\dots , $6$, appearing in that order from bottom to top of the plot. For heat fluxes moments of the same order, the lower lines are for $m = 0$, while the upper lines are for $n = 0$. For (d), heat fluxes moments are shown for $n+m = 2$, $3$, and $6$, appearing in that order from bottom to top of the plot. For heat fluxes moments of the same order, the lower lines are for $n = 0$, while the upper lines are for $m = 0$. Solid lines are the values from the DNS, while squares represent the values from the scaling law. The wall and centre of the channel are swapped, so the centre line is at $x_2/h = 0$, while the wall is at $x_2/h = 1$. Colours as in table \ref{sims}.}
\label{mom_fit_low}
\end{figure}

To analyse the influence of the Prandtl number, we compare the scaling between a high Prandtl number of $4$ and a very low one of $0.01$ in figure \ref{mom_fit_low}. In the case of the scaling of the temperature moments, in figures \ref{t_mom_fit_500_4} and \ref{t_mom_fit_500_001}, for cases $Pr = 4$ and $0.01$, respectively, there is a noticeable difference. While for $Pr = 4$ the scaling (\ref{scaling_law}) represents the DNS data with high accuracy, errors are lower than $0.01 \%$, for the case of $Pr = 0.01$, the deviation is high and clearly visible in figure \ref{t_mom_fit_500_001}. The reason for this error comes from the assumption in section \textsection\ref{sec3} of zero heat conduction in the symmetry analysis, which, obviously, for $Pr = 0.01$ is not true. The high diffusivity for very low Prandtl numbers affects the temperature field in a deeper region away from the wall, and the temperature moments are no longer parallel for $x_2/h$ approximately greater than $0.2$. In the same manner, the scaling of mixed moments is no longer correct for very low Prandtl numbers. While in figure \ref{ut_mom_fit_500_4}, the scaling is again very accurate for $Pr = 4$, in figure \ref{ut_mom_fit_500_001}, a similar failure, as obtained in figure \ref{t_mom_fit_500_001}, appears in the scaling for $Pr = 0.01$. Note that in figure \ref{ut_mom_fit_500_001}, only moments for $n+m = 2$, $4$, and $6$ are plotted for clarity of the figure.

One important point to note is that the functional structure of the invariant solution (\ref{inv_sol}) that leads to the scaling law (\ref{scaling_law}) does not change for vanishing or not vanishing viscosity/diffusivity, in the sense that the same number of parameter appears in the exponent of the scaling law. Therefore, one should expect the scaling law to be correct also for these cases. This is true for a limited region of the centre of the channel. In the case of figures \ref{t_mom_fit_500_001} and \ref{ut_mom_fit_500_001}, if one tries to use the scaling law only for in the region of $y^*<0.1$ away from the centre, instead of $y^*<0.75$, as was done in figure \ref{mom_fit_low}, then a perfect matching with the DNS data will be obtained, even for the lowest Prandtl number cases. As mentioned above, this happens because the viscous/diffusive effects appear so deep away from the wall. Effectively, as shown in Ref.~\onlinecite{alc18}, for such low Reynolds and Prandtl numbers, one cannot even see the emergence of a logarithmic layer, so a bad scaling is also expected. Also, remark that the coefficient $\sigma_1$ and $\sigma_2$ approach to a value of $2$ as the region where the fitting is done is reduced.

Note here that although a $Re_\tau = 500$ is a low Reynolds number and one may expect the assumption of vanishing viscosity to fail, the velocity field is still turbulent and in the centre of the channel, the mentioned assumption is true. However, $Pr = 0.01$ produces a much less turbulent temperature field, even also laminar \cite{alc18}, and for this reason, the assumption of vanishing heat conduction is not true for such a low Prandtl number, or at least is only true in a very central region of the channel (less than $10\%$), where the temperature moments are parallel in figure \ref{t_mom_fit_500_001}. In addition, the key parameter is actually not the Prandtl number by itself, but the friction P\'eclet number, defined as $Pe_\tau = Re_\tau Pr$. Therefore, in our plots, we are comparing a $Re_\tau = 500$ with $Pr = 4$ and $0.01$, which means $Pe_\tau = 2000$ compared with $Pe_\tau = 5$, explaining the errors in the scaling for such a low Prandtl or friction P\'eclet numbers. 

It is important to analyse the values of the different exponential parameters of the scaling law (\ref{scaling_law}) and see the influence of each symmetry on the final scaling. As it was shown in equation (\ref{scaling_law}), the exponent of the power-law is formed by a constant term, $2\sigma_1 - \sigma_2$, that comes from the statistical scaling symmetry of the moments, $\overline{T}_{Ss}$ (\ref{sym_MPC_sca}). A second term that scales with $n$, i.e. $n(\sigma_2 - \sigma_1)$, emerged from the classical scaling symmetries of space and time, $\overline{T}_{Sx}$ (\ref{sym_MPC_sca_sp}) and $\overline{T}_{St}$ (\ref{sym_MPC_sca_t}), respectively. A third term that scales with $m$, i.e. $m\sigma_\Theta$, has its roots in the scaling symmetry of the temperature, $\overline{T}_{S\Theta}$ (\ref{sym_MPC_sca_theta}). However, as can be clearly seen in figure \ref{mom_fit_high}, all moments have a more or less constant slope in the log-log plot, which translates into a very weak dependence on $n$ and $m$. In other words, the values of $\sigma_1$ and $\sigma_2$ are very similar, and $\sigma_\Theta$ is small compared with the value of $2\sigma_1 - \sigma_2$. This, in turn, implies that scaling of space and time has almost no influence in the centre of the channel, and the statistical scaling of moments is dominant, which makes sense, since, as it was said before, it is a measure of intermittency. Scaling independent of the dimensions of space and time is called anomalous scaling and has its origin in the intermittency symmetry.

\begin{figure}
\centering
\begin{subfigure}[b] {0.49\textwidth}
\includegraphics[width=0.99\textwidth]{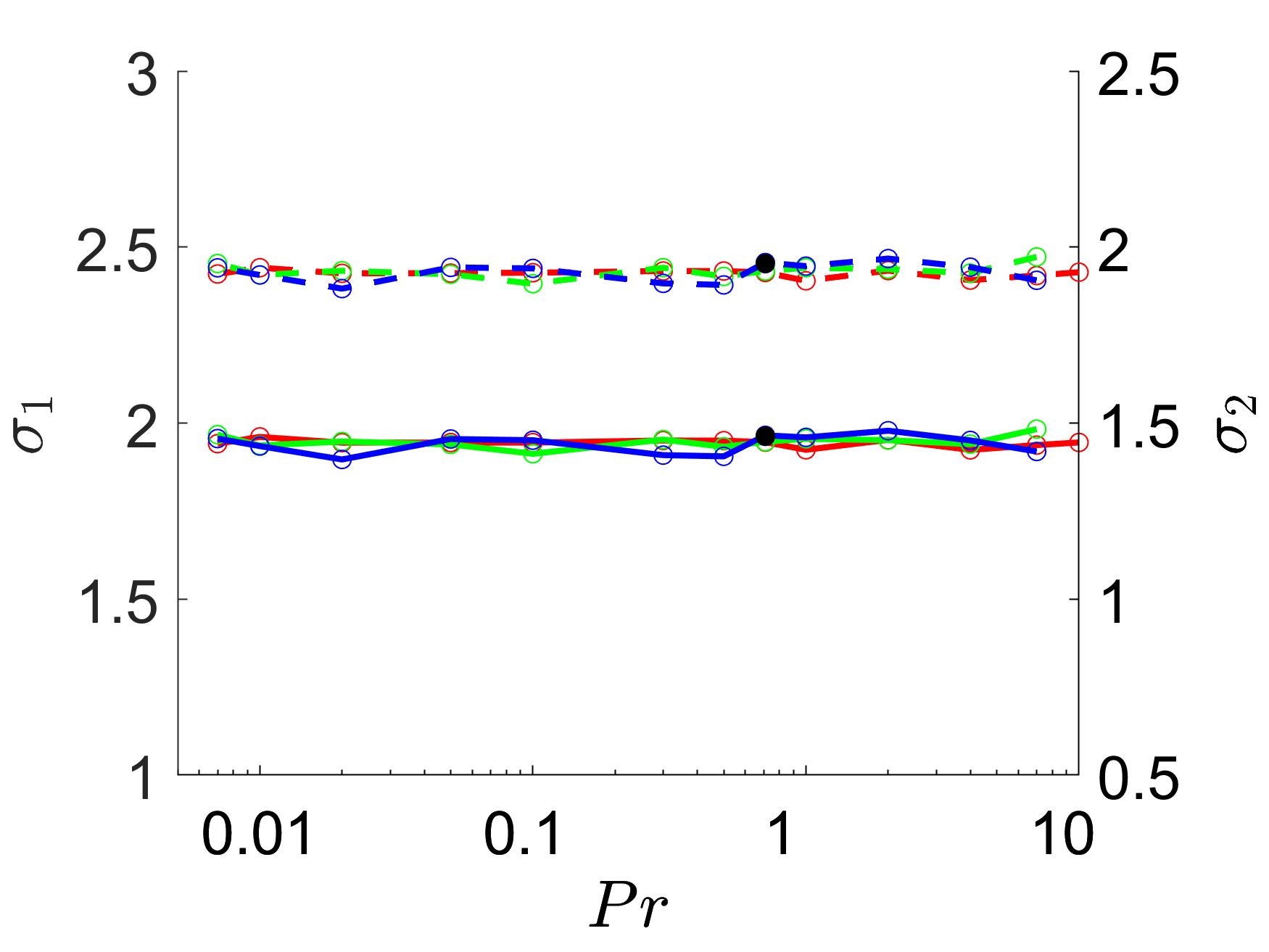}
	\caption{}
	\label{sigma_1_2_Pr}
\end{subfigure}
\centering
\begin{subfigure}[b] {0.49\textwidth}
\includegraphics[width=0.99\textwidth]{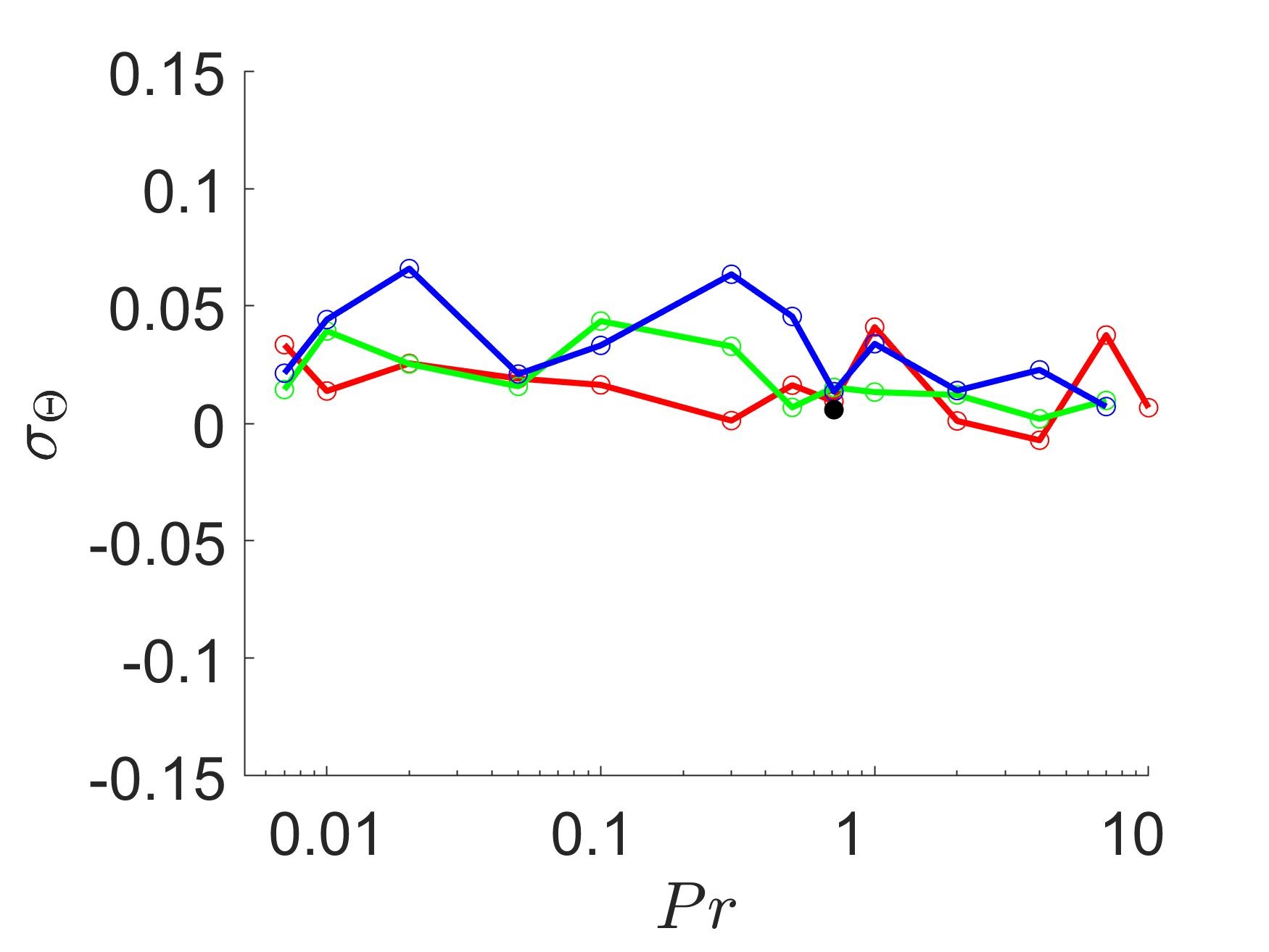}
	\caption{}
	\label{sigma_t}
\end{subfigure}
\caption{Values of (a) $\sigma_1$, solid lines left axis, and $\sigma_2$, dashed lines right axis, and (b) $\sigma_\Theta$. Colours as in table \ref{sims}. Note that black circles at $\Pran = 0.71$ represent the value for the single simulation at $\Rey_\tau = 5000$.}
\label{sigmas}
\end{figure}

Figure \ref{sigma_1_2_Pr} presents the values of $\sigma_1$ and $\sigma_2$ for all the DNS simulations, while the values of $\sigma_\Theta$ are shown in figure \ref{sigma_t}. The values of $\sigma_1$ are almost the same as $\sigma_2$ (note that the left and right axes are shifted for better visualization), which confirms that the symmetries of scaling in space and time have almost no influence in the centre of the channel. Similarly, the scaling symmetry of temperature has barely any influence, since $\sigma_\Theta \ll 2\sigma_1 - \sigma_2$. This last term, $2\sigma_1 - \sigma_2$, is indeed the only dominant term in the exponent of the scaling law (\ref{scaling_law}), with a value slightly below $2$, confirming that the symmetry of scaling of moments, $\overline{T}_{Ss}$ (\ref{sym_MPC_sca}), is dominant in the centre of the channel. Further, we observe that the parameters, $\sigma_1$, $\sigma_2$ and $\sigma_\Theta$, in the investigated ranges, are largely independent of $\Rey_\tau$ and $\Pran$. Small differences are due to small numerical errors or noise in the fitting.

\begin{figure}
\centering
\begin{subfigure}[b] {0.49\textwidth}
\includegraphics[width=0.99\textwidth]{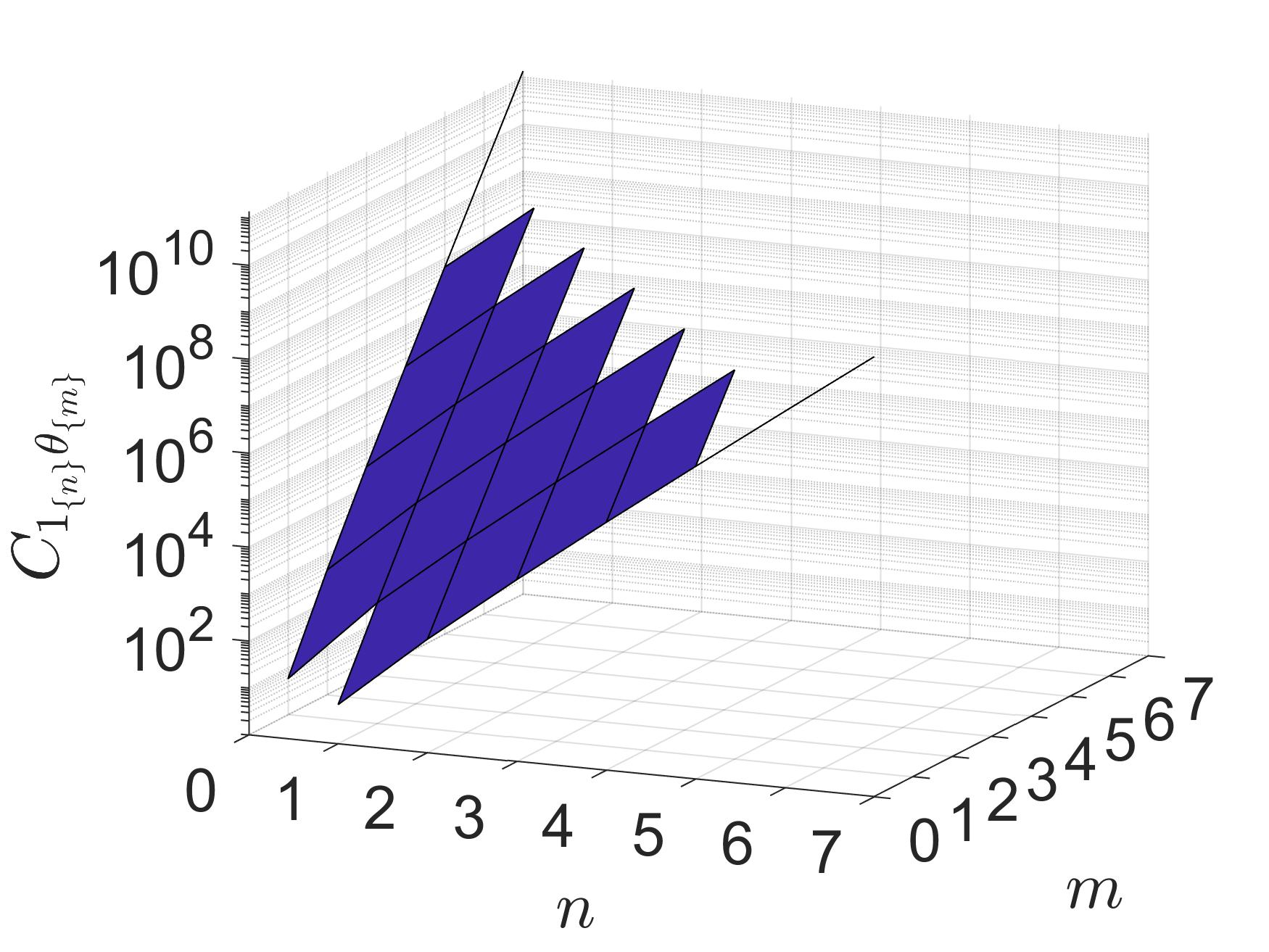}
	\caption{}
	\label{C_500_4}
\end{subfigure}
\centering
\begin{subfigure}[b] {0.49\textwidth}
\includegraphics[width=0.99\textwidth]{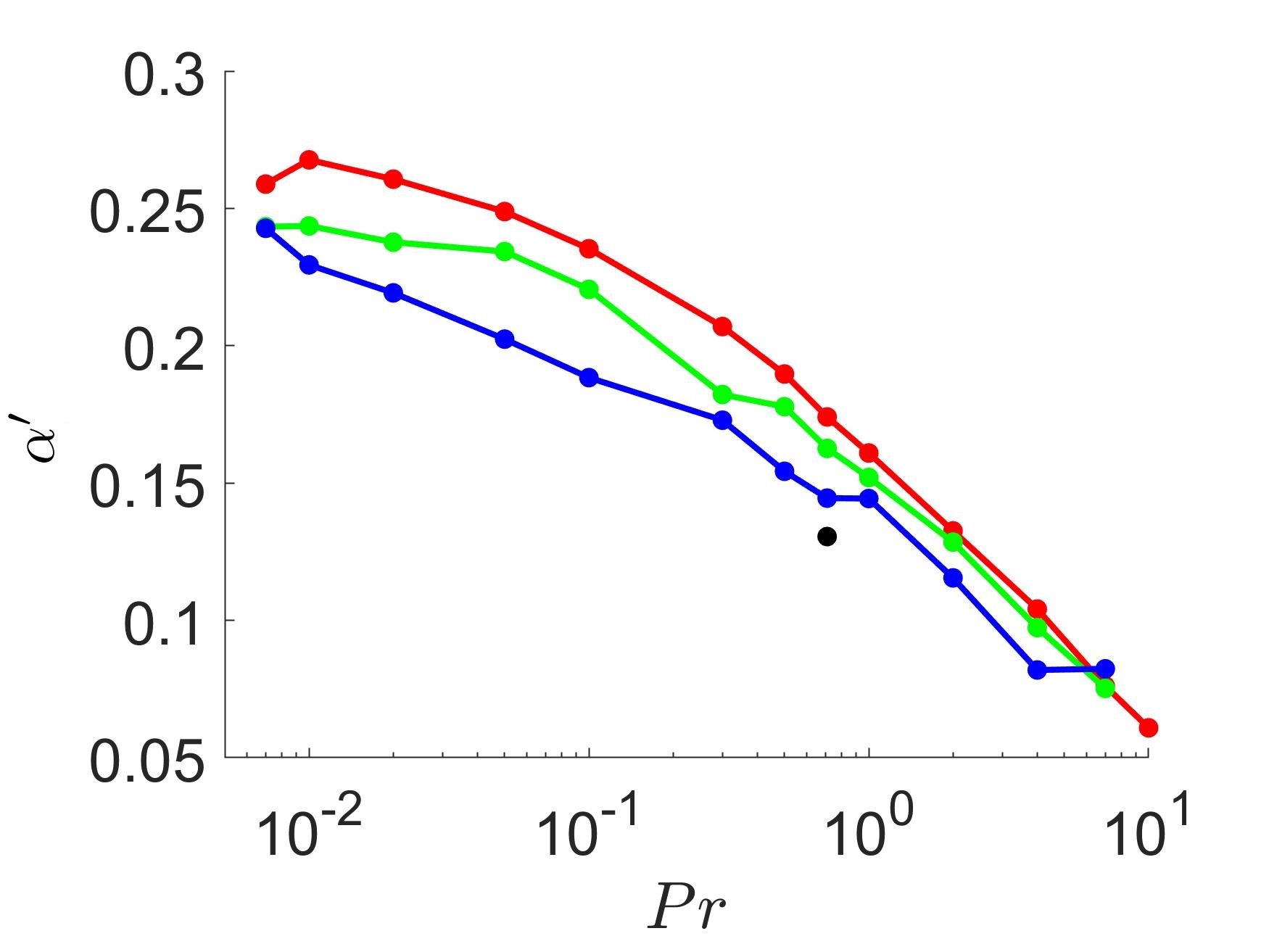}
	\caption{}
	\label{alpha_1t}
\end{subfigure}
\centering
\begin{subfigure}[b] {0.49\textwidth}
\includegraphics[width=0.99\textwidth]{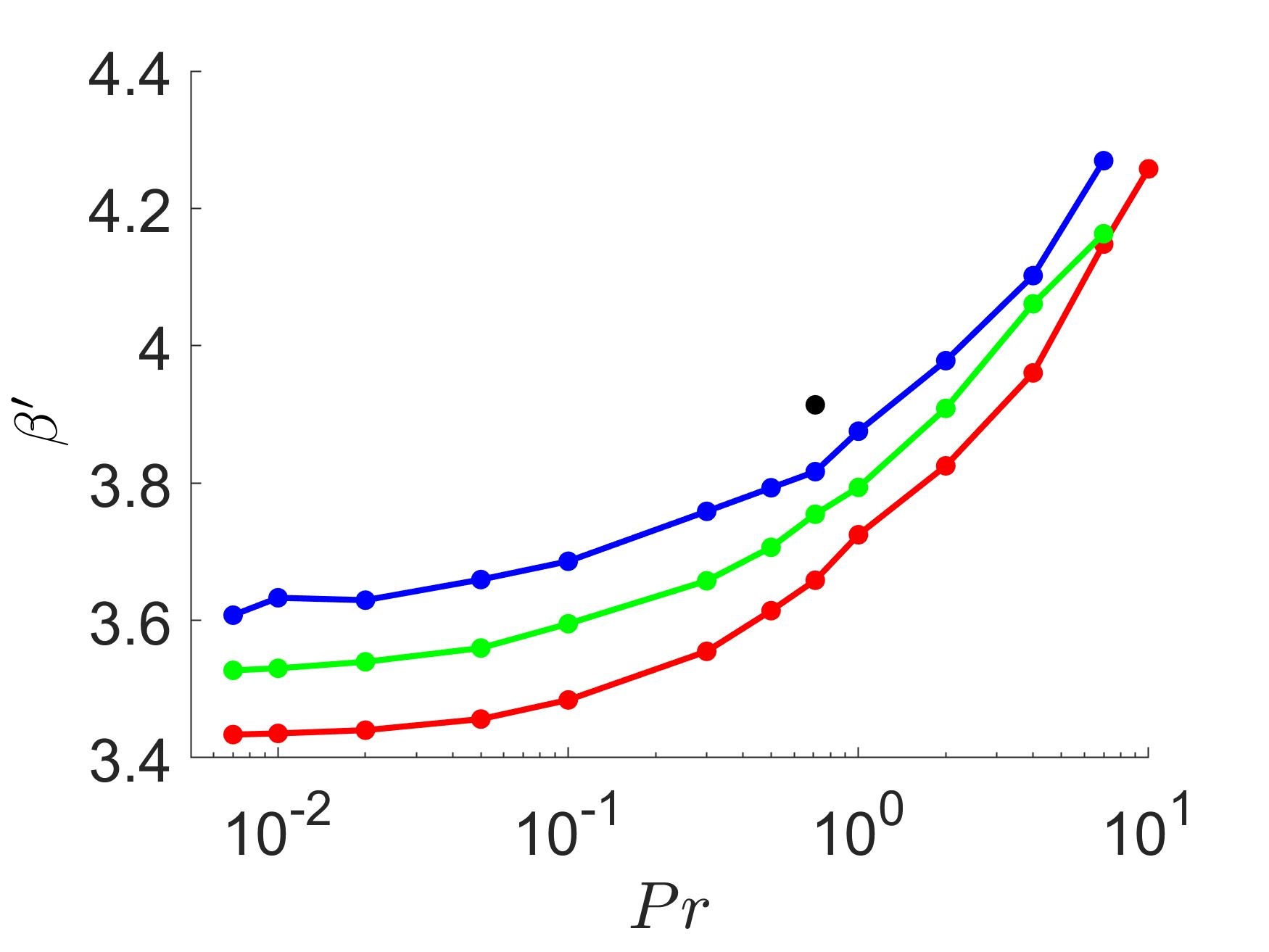}
	\caption{}
	\label{beta_1}
\end{subfigure}
\centering
\begin{subfigure}[b] {0.49\textwidth}
\includegraphics[width=0.99\textwidth]{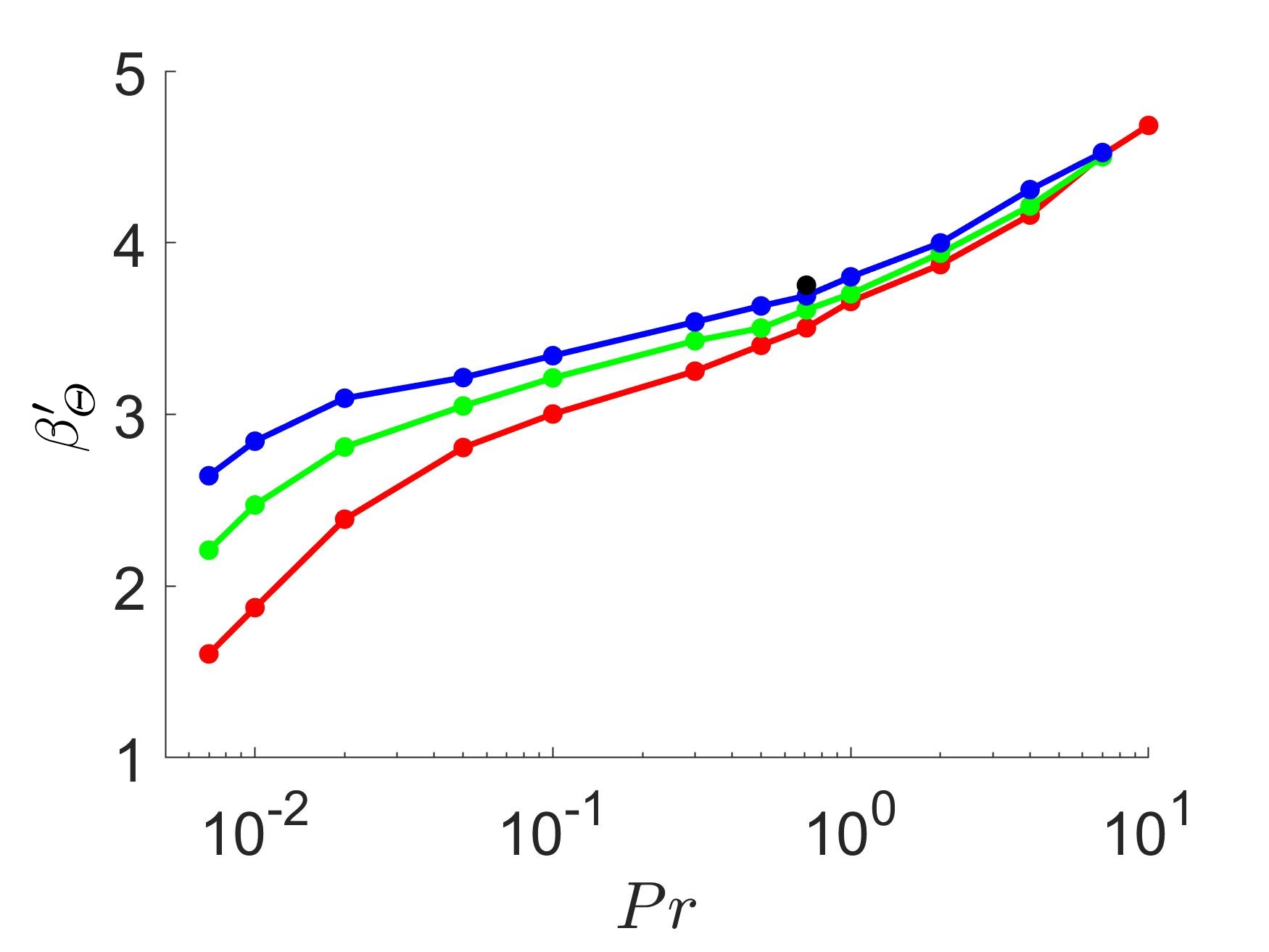}
	\caption{}
	\label{beta_t}
\end{subfigure}
\caption{(a) Values of $C_{1_{\{n\}}\Theta_{\{m\}}}$ for $\Rey_\tau = 500$ and $\Pran = 4$. Parameters from equation (\ref{prefactor}): (b) $\alpha'$, (c) $\beta'$ and (d) $\beta'_\Theta$. Colours as in table \ref{sims}. Note that black points at $\Pran = 0.71$ in (b), (c), and (d) represent the values for the single simulation at $\Rey_\tau = 5000$.}
\label{parameters_C}
\end{figure}

The second important part of the scaling law (\ref{scaling_law}) is the prefactor $C'_{nm}$. Figure \ref{C_500_4} shows the values of $C'_{nm}$ for $\Rey_\tau = 500$ and $\Pran = 4$. An almost perfect plane is observed in the vertical log-scaling plot, which confirms that $C'_{nm}$ is an exponential function in $n$ and $m$, and further verifies that the constants of integration $c'_{nm}$ in equation (\ref{origin_prefactor}) are independent of $n$ and $m$.

In figures \ref{alpha_1t}, \ref{beta_1} and \ref{beta_t}, the values of $\alpha'$, $\beta'$ and $\beta'_\Theta$ for all the present simulations are shown, respectively. To calculate them, only the constants of integration of the moments up to order $2$ have been used, i.e., $C'_{10}$, $C'_{20}$, $C'_{01}$, $C'_{02}$ and $C'_{11}$. With these five values, a fit of the parameters $\alpha'$, $\beta'$, and $\beta'_\Theta$ have been done minimizing again as in equation (\ref{fit}).

Interestingly enough, and other than the parameters in the exponent $\sigma_1$, $\sigma_2$ and $\sigma_\Theta$, the coefficients $\alpha'$, $\beta'$ and $\beta'_\Theta$ are not independent of $\Rey_\tau$ and $\Pran$. We want to point out that for high friction P\'eclet numbers, the values of $\alpha'$, $\beta'$ and $\beta'_\Theta$ seem to be independent of the friction Reynolds number, which may be related to a more realistic assumption of the zero viscosity and heat conduction. However, this is just a point to be investigated. From the theory developed in section \textsection\ref{sec3}, the dependency on $\Rey_\tau$ or $\Pran$ is not apparent but goes beyond the scope of the theory in its present form. Here we limited the study to confirm that the scaling law (\ref{scaling_law}) can represent the behaviour of the arbitrary moments obtained from the DNS data, including exponential scaling of $C'_{nm}$ with $n$ and $m$.

\section{Conclusions}

A new set of turbulent scaling laws for arbitrary moments of the streamwise velocity, temperature, and high order moments of both in a turbulent channel flow has been obtained using the symmetry-based turbulence theory.  These scaling laws apply to incompressible flows driven by a pressure difference and with a passive scalar. For the derivation of the scaling laws, we had to assume vanishing viscosity and diffusion, i.e., $\Rey_\tau \rightarrow \infty$ and $\Pran > 1$, which holds in the central region of the channel, and they are finally cast as deficit laws.

The deficit form of the arbitrary moments in the wall-normal direction can be represented as power functions, where the exponent is determined by the order of the moments and three different parameters that emerged from four different scaling symmetries ($\sigma_1$, $\sigma_2$ and $\sigma_\Theta$). Besides the classical symmetries of the Navier-Stokes and energy equations, we employed statistical symmetries of the multi-point correlation equations, which were the key to obtaining a constant exponent of the power-law scaling function that can accurately represent the DNS data. Instead of the usual fluctuation approach as the basis for the MPC equations, which yields a non-linear system of equations, we presently employ the instantaneous approach, which results in a linear system of equations. The statistical symmetries are trivially displayed in this representation as scaling and translation of moments. This statistical scaling of moments represents a measure of intermittency. It appears as the dominant term in the exponent of all moments as the constant $2\sigma_1 - \sigma_2$, which is independent of the moment order.

The scaling laws have been validated with data from different DNS at different Reynolds and Prandtl numbers. The accuracy of the scaling laws to represent the data is remarkable, especially for high P\'eclet numbers. For cases with low P\'eclet numbers, the centre of the channel gets influenced by viscosity and heat conduction, and the assumption of $Re_\tau \rightarrow \infty$ and $Pr > 1$ no longer holds, which entails a significant deviation from the theoretical scaling of the moments in the centre of the channel.

The exponential prefactor in $n$ and $m$ in equation (\ref{prefactor}) has been obtained by the observation that the constants of integration $c'_{nm}$ in \ref{origin_prefactor} are indeed constant and independent of $n$ and $m$. So far, no justification based on first principles can be given for this, though we speculate that the Probability Density Function (PDF) contains deeper information on this. Therefore, we presently follow the idea of deriving invariant solutions to the PDF equations.

One point of this theory to be explored is the derivation of scaling laws for other important statistics such as fluctuating quantities, cross velocities, the wall-normal and spanwise heat fluxes, and their high-order moments. So far, with the symmetries obtained in this work, it was impossible to properly describe these statistics. However, as mentioned before, further symmetries can be obtained from the MPC equations or from other equations that describe turbulence, such as the PDF equations. This is left as future work which is already being investigated, but we would like to point out that our method is able to obtain scaling laws following only strong mathematical arguments, removing lucky curve fitting.

\section*{Acknowledgments}
This work was supported by PID2021-128676OB-I00 of MINECO/FEDER. FAA is partially funded by GVA/FEDER project ACIF2018. The computations of the new simulations were made possible by a generous grant of computing time from the Barcelona Supercomputing Centre, reference AECT-2020-1-0024. MO expresses his gratitude for the partial support of the German Research Foundation (DFG) within the project OB 96/48-1. We are grateful to Mr Jonathan Laux for providing us with the scripts to do the fittings of the scaling laws with the data.
Declaration of Interests. The authors declare that they have no known competing financial interests or personal relationships that could have appeared to influence the work reported in this paper.

\appendix
\section{}\label{appA}

In this appendix, we present a step-by-step derivation of the MPC equation of the mixed moments. The derivation of the MPC equations of velocity and thermal energy are just two specific cases of the general MPC equations.

Starting with equations (\ref{momentum}) and (\ref{energy}) we perform the following operation to obtain the MPC equation of order $n+m$ of the mixed moments
\begin{align}
&\overline{\mathcal{M}_{i_{(1)}}(\boldsymbol{x}_{(1)}) U_{i_{(2)}}(\boldsymbol{x}_{(2)}) \dots  U_{i_{(n)}}(\boldsymbol{x}_{(n)})\Theta(\boldsymbol{x}_{(n+1)}) \dots  \Theta(\boldsymbol{x}_{(n+m)})} \nonumber\\
& + \overline{U_{i_{(1)}}(\boldsymbol{x}_{(1)}) \mathcal{M}_{i_{(2)}}(\boldsymbol{x}_{(2)}) U_{i_{(3)}}(\boldsymbol{x}_{(3)}) \dots  U_{i_{(n)}}(\boldsymbol{x}_{(n)}) \Theta(\boldsymbol{x}_{(n+1)}) \dots  \Theta(\boldsymbol{x}_{(n+m)})} \nonumber\\
& +\quad \dots  \nonumber\\
& + \overline{U_{i_{(1)}}(\boldsymbol{x}_{(1)}) \dots  U_{i_{(n-1)}}(\boldsymbol{x}_{(n-1)}) \mathcal{M}_{i_{(n)}}(\boldsymbol{x}_{(n)}) \Theta(\boldsymbol{x}_{(n+1)}) \dots  \Theta(\boldsymbol{x}_{(n+m)})} \nonumber\\
& + \overline{U_{i_{(1)}}(\boldsymbol{x}_{(1)}) \dots  U_{i_{(n)}}(\boldsymbol{x}_{(n)}) \mathcal{E}(\boldsymbol{x}_{(n+1)}) \Theta(\boldsymbol{x}_{(n+2)}) \dots  \Theta(\boldsymbol{x}_{(n+m)})} \nonumber\\
& + \overline{U_{i_{(1)}}(\boldsymbol{x}_{(1)}) \dots  U_{i_{(n)}}(\boldsymbol{x}_{(n)}) \Theta(\boldsymbol{x}_{(n+1)}) \mathcal{E}(\boldsymbol{x}_{(n+2)}) \Theta(\boldsymbol{x}_{(n+3)}) \dots  \Theta(\boldsymbol{x}_{(n+m)})} \nonumber\\
& +\quad \dots  \nonumber\\
& + \overline{U_{i_{(1)}}(\boldsymbol{x}_{(1)}) \dots  U_{i_{(n)}}(\boldsymbol{x}_{(n)}) \Theta(\boldsymbol{x}_{(n+1)}) \dots  \Theta(\boldsymbol{x}_{(n+m-1)})}  \mathcal{E}(\boldsymbol{x}_{(n+m)}) = \nonumber\\
= & \sum_{a=1}^n\overline{\mathcal{M}_{i_{(a)}}(\boldsymbol{x}_{(a)})\prod_{c=1,c\neq a}^n U_{i_{(c)}}(\boldsymbol{x}_{(c)})\prod_{d=n+1}^{n+m} \Theta(\boldsymbol{x}_{(d)})} \nonumber\\
& + \sum_{b=n+1}^{n+m}\overline{\mathcal{E}(\boldsymbol{x}_{(b)})\prod_{c=1}^n U_{i_{(c)}}(\boldsymbol{x}_{(c)})\prod_{d=n+1,d\neq c}^{n+m} \Theta(\boldsymbol{x}_{(d)})} = 0,
\label{MPC_HF_start}
\end{align}
where $U_{i_{(l)}}$ and $\boldsymbol{x}_{(l)}$ are the velocity and the different points where the equations and the variables are applied, for $i_{(l)} = 1$, $2$, $3$; and $l = 1$, $2$,\dots ,$n+m$ ($l$ can be $a$ or $b$). Introducing the momentum and energy equations, (\ref{momentum}) and (\ref{energy}), into (\ref{MPC_HF_start}), we obtain
\begin{align}
& \sum_{a=1}^n\overline{\frac{\partial U_{i_{(a)}}(\boldsymbol{x}_{(a)})}{\partial t}\prod_{c=1,c\neq a}^n U_{i_{(c)}}(\boldsymbol{x}_{(c)})\prod_{d=n+1}^{n+m} \Theta(\boldsymbol{x}_{(d)})} \nonumber\\
& + \sum_{a=1}^n\overline{U_k(\boldsymbol{x}_{(a)})\frac{\partial U_{i_{(a)}}(\boldsymbol{x}_{(a)})}{\partial x_{k_{(a)}}}\prod_{c=1,c\neq a}^n U_{i_{(c)}}(\boldsymbol{x}_{(c)})\prod_{d=n+1}^{n+m} \Theta(\boldsymbol{x}_{(d)})} \nonumber\\
& + \sum_{a=1}^n\overline{\frac{\partial P(\boldsymbol{x}_{(a)})}{\partial x_{i_{(a)}}}\prod_{c=1,c\neq a}^n U_{i_{(c)}}(\boldsymbol{x}_{(c)})\prod_{d=n+1}^{n+m} \Theta(\boldsymbol{x}_{(d)})} \nonumber\\
&- \frac{1}{\Rey_\tau}\sum_{a=1}^n\overline{\frac{\partial^2 U_{i_{(a)}}(\boldsymbol{x}_{(a)})}{\partial x_{k_{(a)}} \partial x_{k_{(a)}}}\prod_{c=1,c\neq a}^n U_{i_{(c)}}(\boldsymbol{x}_{(c)})\prod_{d=n+1}^{n+m} \Theta(\boldsymbol{x}_{(d)})} \nonumber\\
& + \sum_{b=n+1}^{n+m}\overline{\frac{\partial \Theta(\boldsymbol{x}_{(b)})}{\partial t}\prod_{c=1}^n U_{i_{(c)}}(\boldsymbol{x}_{(c)})\prod_{d=n+1,d\neq c}^{n+m} \Theta(\boldsymbol{x}_{(d)})}, \nonumber\\
& + \sum_{b=n+1}^{n+m}\overline{U_k(\boldsymbol{x}_{(b)})\frac{\partial \Theta(\boldsymbol{x}_{(b)})}{\partial x_{k_{(b)}}}\prod_{c=1}^n U_{i_{(c)}}(\boldsymbol{x}_{(c)})\prod_{d=n+1,d\neq c}^{n+m} \Theta(\boldsymbol{x}_{(d)})}, \nonumber\\
& - \frac{1}{\Pen_\tau} \sum_{b=n+1}^{n+m}\overline{\frac{\partial^2 \Theta(\boldsymbol{x}_{(b)})}{\partial x_{k_{(b)}} \partial x_{k_{(b)}}}\prod_{c=1}^n U_{i_{(c)}}(\boldsymbol{x}_{(c)})\prod_{d=n+1,d\neq c}^{n+m} \Theta(\boldsymbol{x}_{(d)})} = 0.
\label{MPC_HF_intro}
\end{align}

At this point, the continuity equation (\ref{continuity}) should be applied to introduce the terms $U_k(\boldsymbol{x}_{l})$ inside the derivatives with respect to $x_{k_{(l)}}$ in the second and sixth lines of equation (\ref{MPC_HF_intro}). Also, the product terms can be introduced in the derivatives with respect to the spatial coordinates, since the points $\boldsymbol{x}_{(a)}$ and $\boldsymbol{x}_{(b)}$ are excluded from the product series. Regarding the temporal derivatives in the first and fifth lines of (\ref{MPC_HF_intro}), the chain rule is applied to reduce it to a single term. Finally, using definitions (\ref{def_HF}), (\ref{def_PP}) and (\ref{def_change}) one can obtain the MPC equation for all mixed moments written in the following way
\begin{align}
& \frac{\partial H_{i_{\{n\}}\Theta_{\{m\}}}}{\partial t} + \nonumber\\
& \sum_{a=1}^n\left(\frac{\partial H_{i_{\{n+1\}}\Theta_{\{m\}}[i_{(n+m+1)}\rightarrow k]}(\boldsymbol{x}_{(n+m+1)}\rightarrow\boldsymbol{x}_{(a)})}{\partial x_{k_{(a)}}} + \frac{\partial I_{i_{\{n-1\}}\Theta_{\{m\}}[a]_{P}}}{\partial x_{i_{(a)}}} - \frac{1}{\Rey_\tau}\frac{\partial^2 H_{i_{\{n\}}\Theta_{\{m\}}}}{\partial x_{k_{(a)}}\partial x_{k_{(a)}}}\right) \nonumber\\
& + \sum_{b=n+1}^{n+m}\left(\frac{\partial H_{i_{\{n+1\}}\Theta_{\{m\}}[i_{(n+m+1)}\rightarrow k]}(\boldsymbol{x}_{(n+m+1)}\rightarrow\boldsymbol{x}_{(b)})}{\partial x_{k_{(b)}}} - \frac{1}{\Pen_\tau}\frac{\partial^2 H_{i_{\{n\}}\Theta_{\{m\}}}}{\partial x_{k_{(b)}}\partial x_{k_{(b)}}}\right) = 0.
\label{MPC_HF_final}
\end{align}

As mentioned before, the MPC equations of the velocity arises if $m = 0$ in (\ref{MPC_HF_final}). Similarly, one can obtain the MPC equations of the temperature by setting $n = 0$ in (\ref{MPC_HF_final}).

\section{}\label{appB}

In this appendix, the Two-Point Correlation (TPC) equations for the velocity, heat fluxes and temperature are given as examples of the MPC equation (\ref{MPC}), in order to make the notation clearer. These equations can be obtained by setting in equation (\ref{MPC}) $n=2$, $1$, $0$ and $m = 0$, $1$, $2$, respectively.
\begin{align}
& \frac{\partial H_{i_{(1)}i_{(2)}}(\boldsymbol{x}_{(1)},\boldsymbol{x}_{(2)})}{\partial t} + \frac{\partial H_{i_{(1)}i_{(2)}k}(\boldsymbol{x}_{(1)},\boldsymbol{x}_{(2)},\boldsymbol{x}_{(1)})}{\partial x_{k_{(1)}}} + \frac{\partial H_{i_{(1)}i_{(2)}k}(\boldsymbol{x}_{(1)},\boldsymbol{x}_{(2)},\boldsymbol{x}_{(2)})}{\partial x_{k_{(2)}}} \nonumber\\
& + \frac{\partial I_{Pi_{(2)}}(\boldsymbol{x}_{(1)},\boldsymbol{x}_{(2)})}{\partial x_{i_{(1)}}} + \frac{\partial I_{i_{(1)}P}(\boldsymbol{x}_{(1)},\boldsymbol{x}_{(2)})}{\partial x_{i_{(2)}}} \nonumber\\
& - \nu\frac{\partial^2 H_{i_{(1)}i_{(2)}}(\boldsymbol{x}_{(1)},\boldsymbol{x}_{(2)})}{\partial x_{k_{(1)}}\partial x_{k_{(1)}}} - \nu\frac{\partial^2 H_{i_{(1)}i_{(2)}}(\boldsymbol{x}_{(1)},\boldsymbol{x}_{(2)})}{\partial x_{k_{(2)}}\partial x_{k_{(2)}}}  = 0,\label{TPC_M}\\
& \frac{\partial H_{i_{(1)}\Theta}(\boldsymbol{x}_{(1)},\boldsymbol{x}_{(2)})}{\partial t} + \frac{\partial H_{i_{(1)}\Theta k}(\boldsymbol{x}_{(1)},\boldsymbol{x}_{(2)},\boldsymbol{x}_{(1)})}{\partial x_{k_{(1)}}} + \frac{\partial H_{i_{(1)}\Theta k}(\boldsymbol{x}_{(1)},\boldsymbol{x}_{(2)},\boldsymbol{x}_{(2)})}{\partial x_{k_{(2)}}} \nonumber\\
& + \frac{\partial I_{P\Theta}(\boldsymbol{x}_{(1)},\boldsymbol{x}_{(2)})}{\partial x_{i_{(1)}}} \nonumber\\
& - \nu\frac{\partial^2 H_{i_{(1)}\Theta}(\boldsymbol{x}_{(1)},\boldsymbol{x}_{(2)})}{\partial x_{k_{(1)}}\partial x_{k_{(1)}}} - \alpha\frac{\partial^2 H_{i_{(1)}\Theta}(\boldsymbol{x}_{(1)},\boldsymbol{x}_{(2)})}{\partial x_{k_{(2)}}\partial x_{k_{(2)}}} = 0, \label{TPC_HF}\\
& \frac{\partial H_{\Theta\Theta}(\boldsymbol{x}_{(1)},\boldsymbol{x}_{(2)})}{\partial t} + \frac{\partial H_{\Theta\Theta k}(\boldsymbol{x}_{(1)},\boldsymbol{x}_{(2)},\boldsymbol{x}_{(1)})}{\partial x_{k_{(1)}}} + \frac{\partial H_{\Theta\Theta k}(\boldsymbol{x}_{(1)},\boldsymbol{x}_{(2)},\boldsymbol{x}_{(2)})}{\partial x_{k_{(2)}}} \nonumber\\
& - \alpha\frac{\partial^2 H_{\Theta\Theta}(\boldsymbol{x}_{(1)},\boldsymbol{x}_{(2)})}{\partial x_{k_{(1)}}\partial x_{k_{(1)}}} - \alpha\frac{\partial^2 H_{\Theta\Theta}(\boldsymbol{x}_{(1)},\boldsymbol{x}_{(2)})}{\partial x_{k_{(2)}}\partial x_{k_{(2)}}} = 0. \label{TPC_E}
\end{align}

\bibliography{turbulence}

\end{document}